%% file: MarkovLimFb_Arxiv.tex
\documentclass[11pt,onecolumn,draftclsnofoot]{IEEEtran}
\usepackage{fancyhdr}
\usepackage{amsmath,epsfig}
\usepackage{threeparttable}
\usepackage{epsf,epsfig}
\usepackage{amsmath}
\usepackage{amssymb}
\usepackage{amsfonts}
\usepackage{algorithmic}
\usepackage{algorithm}
\usepackage[noadjust]{cite}
\usepackage{dsfont}
\usepackage{subfigure}

\newcommand{\bpi}{\boldsymbol{\pi}}

\begin{document}
\include{header}
\title{\huge \setlength{\baselineskip}{30pt} Limited Feedback Beamforming Over Temporally-Correlated Channels}
\author{\large \setlength{\baselineskip}{15pt}Kaibin Huang, Robert W. Heath, Jr., and Jeffrey G. Andrews
\thanks{\setlength{\baselineskip}{14pt} K. Huang, R. W. Heath, Jr.,  and J. G. Andrews  are with Wireless Networking and Communications Group, Department of Electrical and Computer Engineering, The University of Texas at Austin, 1 University Station C0803, Austin, TX 78712. Email: khuang@mail.utexas.edu, \{rheath, jandrews\}@ece.utexas.edu. K. Huang is the recipient of the University Continuing Fellowship from The University of Texas. This work is funded by the DARPA IT-MANET program under the grant W911NF-07-1-0028, and the National Science Foundation under the grant CCF-514194. The results in this paper were presented in part at the IEEE Int. Conf. Acoust., Speech and Sig. Proc., May 2006, and the IEEE GLOBECOM, Nov. 2006.}}
\markboth{Submitted to IEEE Transactions on Signal Processing, Revised on \today}{}\maketitle
\vspace{-50pt}

\begin{abstract}\setlength{\baselineskip}{15pt}
Feedback of quantized channel state information (CSI), called \emph{limited feedback}, enables transmit beamforming in multiple-input-multiple-output (MIMO) wireless systems with a small amount of overhead. Due to its efficiency, beamforming with limited feedback has been adopted in several wireless communication standards. Prior work on limited feedback commonly adopts the block fading channel model where temporal correlation in wireless channels is neglected. This paper considers temporally-correlated channels and designs single-user transmit beamforming with limited feedback. Analytical results concerning CSI feedback are derived by modeling quantized CSI as a first-order finite-state Markov chain. These results include the source bit rate generated by time-varying quantized CSI, the required bit rate for a CSI feedback channel, and the effect of feedback delay. In particular, based on the theory of Markov chain convergence rate, feedback delay is proved to reduce the throughput gain due to CSI feedback at least exponentially. Furthermore, an algorithm is proposed for CSI feedback compression in time. Combining the results in this work leads to a new method for designing limited feedback beamforming as demonstrated by a design example.
\end{abstract}


\section{Introduction}
\label{Section:Introd}
For a multiple-input-multiple-out (MIMO) communication system, transmit beamforming alleviates the negative effect of channel fading by exploiting spatial diversity, and thereby increases system throughput. Typically, transmit beamforming requires feedback of channel state information (CSI) from a receiver to a transmitter. Such CSI feedback can potentially incur excessive overhead due to the multiplicity of channel coefficients. This motivates designing highly-efficient CSI quantization algorithms based on  communication measures such as information capacity. These algorithms for both beamforming and other MIMO techniques form an active field called \emph{limited feedback} (see e.g. \cite{LoveHeathBook} and the references therein). This paper focuses on limited feedback beamforming systems over temporally correlated channels. It  addresses a set of open issues including the information rate inherent in CSI, the required bit rate for a CSI feedback  channel, the effect of feedback delay on throughput, and feedback CSI compression. This paper provides a systematic method of designing the CSI feedback link of a limited feedback beamforming system.

\subsection{Prior Work and Motivation}\label{Section:PriorWork}
For simplicity, most prior work on limited feedback adopts the  \emph{block fading} channel model, where each channel realization remains constant in one block and different realizations are independent \cite{GoldsmithBook:WirelessComm:05}. Using this model, designing limited feedback reduces to a vector quantization problem \cite{GerGra:VectQuanSignComp:92}. Different methods for quantizing CSI have been developed such as line packing \cite{LovHeaETAL:GrasBeamMultMult:Oct:03,MukSabETAL:BeamFiniRateFeed:Oct:03}, combined parametrization and scalar quantization \cite{Roh:EffFbMIMOParameter:2007}, subspace interpolation \cite{ChoiMondal:InterpPrecodSpaMuxMIMOOFDMLimFb:2006}, and Lloyd's algorithm \cite{Xia:AchieveWelchBound:05}. Furthermore, different types of limited feedback systems have been investigated, including beamforming \cite{LovHeaETAL:GrasBeamMultMult:Oct:03, MukSabETAL:BeamFiniRateFeed:Oct:03}, precoded orthogonal space-time block codes \cite{LoveHeath:LimitedFeedbackPrecodOSTBC:05}, precoded spatial multiplexing \cite{LoveHeath:LimitedFeedbackPrecodSpatialMultiplex:05},
and multiuser downlink \cite{Gesbert:ShiftMIMOParadigm:2007}. Prior work focuses on  analyzing and minimizing CSI inaccuracy due to quantization. The existing results do not account for how channel coherence time influences the information rate generated by time-varying CSI,  and the bit rate that a feedback channel must support. These issues are addressed in this paper.

For practical systems, the block fading assumption in prior work is pessimistic since channel temporal correlation often exists and can be used for compressing feedback CSI by incremental feedback. Algorithms for feedback compression have been proposed in \cite{Roh:EffFbMIMOParameter:2007, Banister03, SimonLeus:FbReductionSMLinearPrecod:2007} that exploit channel temporal correlation. In \cite{Roh:EffFbMIMOParameter:2007}, a MIMO channel is parameterized and the feedback of each parameter is compressed to be one bit. Nevertheless, the multiplicity of the channel parameters compromises the feedback efficiency. In \cite{Banister03}, the feedback CSI is compressed to be one bit but requires the transmitter to periodically broadcast channel subspace matrices. Building on the preliminary results of the current work in \cite{Huang:MarkovModelLimitFb:ICASSP06}, variable-length source codes such as a Huffman code \cite{GallagerBook:InfoTheoReliableComm:68} are applied for CSI feedback compression in \cite{SimonLeus:FbReductionSMLinearPrecod:2007}. Despite minimizing the average CSI source bit rate, this approach may not suit practical applications where CSI feedback consists of fixed-length bit blocks \cite{3GPP-LTE, IEEE802-16e}.

In a practical system with CSI feedback, feedback delay exists due to sources such as signal processing, propagation and channel-access protocols. The negative effects of CSI feedback delay on bit-error rate or information capacity have been observed in the literature of MIMO communication
\cite{Isu:FiniteRateFbCorrMISOEstmErrFbDly:2007, Ting:MarkovKronMIMO:2006, Au:AdaptModMultiantTxChannelMeanFb:2004, Nguyen:CapMIMOFbDelay:2004, DuLi:PerformLossFbDelayMIMO:2004, KuoSmith:SelfInterfMIMOFbDelay:2006, Kobayashi:MIMOFbDelay:2006}. Specifically, this delay is found to decrease received signal power and cause interference between spatial data streams. The negative impact of CSI feedback delay can be alleviated to some extent by channel prediction  \cite{Kobayashi:MIMOFbDelay:2006, DuelHallen:FadingChanPredictAdaptTx:2007}. Despite its importance for designing MIMO systems, there exists no simple relationship between CSI feedback delay and throughput. This motivates the current work on deriving such a relationship in the context of limited feedback beamforming. The results of this work have been validated by measurement data from a MIMO prototype over an indoor channel
\cite{Daniels07:QuantizedFbReductionSMLinearPrecod:2007}.

The main approach of this work is to model quantized CSI as a finite-state Markov chain (FSMC), which allows the use of Markov chain theory as an analytical tool. A similar approach is common in modeling single-input-single-output (SISO) fading  channels \cite{Wang:FinStatMarkovChan:95, Pimentel:FiniteMarkovModelCorrRicianFading:2004, Tan:MarkovRayleighFading:2000, Zhang:MarkovModelRayleighFading:1999, Babich:GenMarkovModelFlatFading:2000}. A FSMC model for fading captures the stochastic feature of wireless channels that is missing in block fading. Furthermore, the FSMC model is simple enough for  allowing tractable performance analysis of communication systems. For these reasons, since it was proposed in \cite{Wang:FinStatMarkovChan:95}, this model has be widely applied in wireless channel modeling, communication and networking. The FSMC models have been proposed for the satellite \cite{Babich1999}, indoor \cite{Chen:TractWlssChan:1998}, Rayleigh fading \cite{Saadani:MarkovRayleighRake:2004, Wang:FinStatMarkovChan:95, Tan:MarkovRayleighFading:2000}, Nakagamni fading \cite{Guan:FSMCCorrFading:1999}, and Rician fading \cite{Pimentel:FiniteMarkovModelCorrRicianFading:2004} channels. The accuracy of FSMC models has been verified based on different criteria, including information capacity \cite{Wang:VerifyFirstOrderMarkovRayleigh:1996, Sadeghi:CapAnalysisMarkovFlatFading:2005}, packet errors \cite{Bischl:PacketErrorNonInterleaveRayleigh:1995}, burst errors \cite{Hueda:InfoTheoTestMarkovBlockErrors:2005}, and autocorrelation functions \cite{Tan:MarkovRayleighFading:2000, Pimentel:FiniteMarkovModelCorrRicianFading:2004}. Due to its accuracy and simplicity, the first-order FSMC channel models have been used in designing and analyzing adaptive video encoding \cite{Galluccio2005}, maximum a posteriori decoding \cite{Turin:MAPDecodErrorBurst:2001}, downlink rate control \cite{Razavilar:OptimRateDelayControlFading:2002}, downlink power control \cite{Buche:ControlMobileCommTimeVarchan:2002}, \emph{automatic repeat request} (ARQ) \cite{Yun:MarkovErrARQFading:2005}, and maximum likelihood detection \cite{LiGoldsmith:MLDetectMarkov:2002}. FSMCs have been also adopted for modeling scalar functions of MIMO channel matrices such as the largest singular value \cite{Ivanis:AdpMIMOMRCImpCSIMarkov:2007}, the condition number \cite{Kuo:MarkovMIMOCondNumbAntSel:2007}, and the Frobenius norm \cite{Abdi:LCRAvFadMIMOFading:2003}. Nevertheless, there exist few results on using FSMC for designing MIMO limited feedback systems despite their popularity.


\subsection{Contributions}\label{Section:Contribution}

This paper presents a set of analytical results useful for understanding and designing MIMO limited feedback beamforming over a temporally-correlated channel. These results assume that the time-varying quantized CSI follows a first-order finite-state Markov chain, called \emph{channel state Markov chain}.  The contributions of this paper are summarized as follows.
\begin{enumerate}

\item Quantized CSI is treated as a Markov data source  and the corresponding information bit rate, called \emph{CSI source bit rate}, is derived in terms of the probabilities of the channel state Markov chain. In particular, the CSI source bit rate is shown to decrease linearly with the probability that the channel state remains unchanged over two consecutive data symbol durations. Note that this probability is larger for  longer channel coherence time and vise versa. The CSI source bit rate provides a measure of the rate of information generated by the temporally-correlated channel.

\item CSI feedback can rely on periodic or random-access \cite{TangHeath:OppFbDLMuDiv:2005} feedback protocols, corresponding to separated or shared feedback channels for different users. Based on the periodic feedback protocol, the bit rate supported by a feedback channel, called \emph{CSI feedback bit rate}, is derived under a feedback outage constraint, where an outage refers to multiple channel-state transitions within one feedback interval. This constraint guarantees that CSI feedback is sufficiently frequent. The derived CSI feedback bit rate is useful for allocating bandwidth for the CSI feedback channel.

\item Define the \emph{feedback throughput gain} as the difference in throughput between the cases of delayed CSI feedback and no feedback. Based on the theory of Markov chain convergence rate, the feedback throughput gain is shown to decrease at least exponentially with feedback delay and inversely with the feedback interval. The exponential rate increases with decreasing channel coherence time and vise versa. This result enables the joint design of the mobility speed, the bandwidth of the feedback channel, and feedback delay sources such as signal processing complexity and the propagation distance, under a constraint on the feedback throughput gain. Moreover generally, the feedback throughput gains are proved to be delay sensitive.

\item Finally, an algorithm is proposed for compressing feedback CSI by exploiting residual temporal correlation in feedback CSI. Note that the temporal correlation in CSI is largely reduced by the periodic feedback protocol. This algorithm compresses feedback CSI by truncating low-probability transitions between states of the channel state Markov chain. Moreover, this algorithm alternates compressed and uncompressed CSI feedback to prevent propagation of CSI errors due to such truncation.


\end{enumerate}


The differences between this manuscript and its conference versions \cite{Huang:MarkovModelLimitFb:ICASSP06, Huang:EffFbDelay:Globecom06, Huang:LimitedFeedbackCompression:Globecom06} are highlighted as follows. First, the CSI feedback bit rate derived in this manuscript targets a periodic feedback protocol under a feedback outage constraint, but that in \cite{Huang:MarkovModelLimitFb:ICASSP06} is based on an aperiodic feedback protocol without such a constraint. Note that the periodic feedback protocol is more suitable for existing communication standards such as 3GPP-LTE \cite{3GPP-LTE}, where a fixed number of bits in each reverse-link data block are allocated for CSI feedback. Second, this manuscript considers both fixed feedback delay and variable delay inherent in the periodic feedback protocol, but  only the former is addressed in \cite{Huang:EffFbDelay:Globecom06}. Moreover, the analysis on the effect of fixed feedback in \cite{Huang:EffFbDelay:Globecom06} has been made more rigorous in this manuscript. Third, the accuracy of the Markov chain model for CSI feedback is verified by simulation in this manuscript but not in \cite{Huang:MarkovModelLimitFb:ICASSP06, Huang:EffFbDelay:Globecom06, Huang:LimitedFeedbackCompression:Globecom06}.  Fourth, the CSI feedback compression algorithm proposed in \cite{Huang:LimitedFeedbackCompression:Globecom06} is improved in this manuscript to avoid propagation of CSI errors due to truncation of low-probability channel-state transitions. Finally, a new design example is included in this manuscript for demonstrating the joint application of different analytical results from this work.

\subsection{Organization and Notation}\label{Section:Organization}
The remainder of this paper is organized as follows. In Section~\ref{Section:System}, a limited feedback beamforming system is described.  Section~\ref{Section:ChanMarkov} presents the definition and construction procedure of the channel state Markov chain. In Section~\ref{Section:CSIRate:Bandwidth}, CSI source and feedback bit rates are derived. Section~\ref{Section:FbDelay} focuses on the relationship between the feedback throughput gain and feedback delay. In Section~\ref{Section:FBCompress}, an algorithm for CSI feedback compression is proposed and analyzed, followed by concluding remarks in Section~\ref{Section:Conclusion}.

{\bf Notation:} Capitalized and small boldface letters denote matrices and vectors, respectively.  The superscript $*$ represents the complex conjugate and transpose matrix operation. The operator $[\cdot]_{\ell}$ gives the $\ell$th component of a matrix. Similarly, $[\cdot]_{\ell m}$ returns the $(\ell, m)$th component of a vector.

\section{System Description} \label{Section:System}
The limited feedback beamforming system illustrated in Fig.~\ref{Fig:System} can be separated into the forward and the CSI feedback links, which are described in the following sub-sections. In this system, all signals are discrete-time and sampled at the sampling rate $1/T$, where $T$ denotes one sample interval. Without loss of generality, $T$ is set equal to one symbol duration. The sample index is denoted by the subscript $n$.

\subsection{Forward Link}\label{Section:System:Forward}
\begin{figure}
\centering
  \includegraphics[width=15cm]{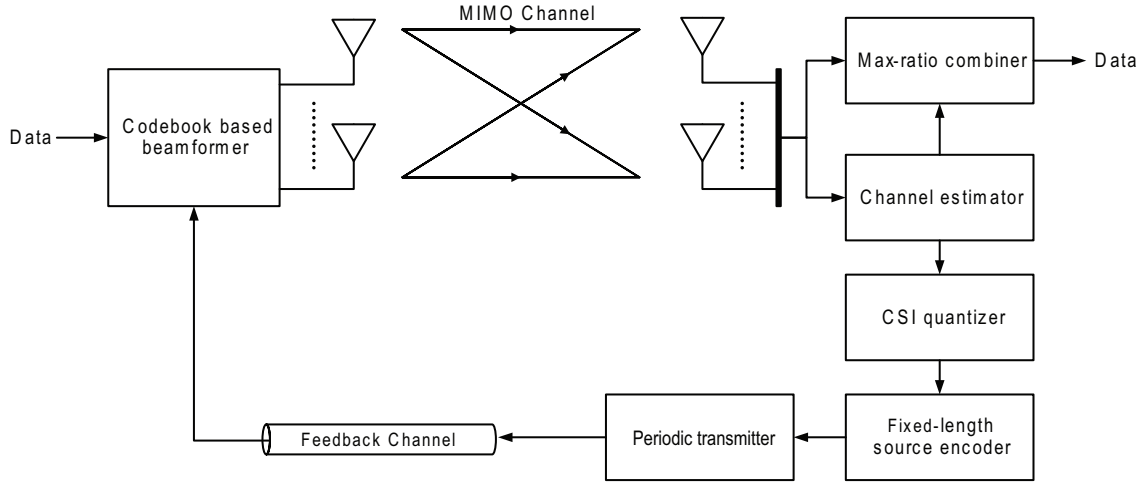}\\
  \caption{Limited feedback beamforming system} \label{Fig:System}
\end{figure}

The forward link refers to the data path in Fig.~\ref{Fig:System} from the input of the beamformer to the output of the maximum-ratio combiner.
The $n$th received data symbol after maximum ratio combining, denoted as $y_n$,  is given as
\begin{equation}\label{Eq:Sys}
    y_n = \sqrt{P}\bw^*_n\bH_n\bff_nx_n + \nu_n
\end{equation}
where $x_n$ and $\nu_n$ are  $\mathcal{CN}(0,1)$ random variables modeling respectively the $n$th transmitted data symbol and the $n$th sample of the additive white Gaussian noise (AWGN) process, and $P$ is the transmit signal-to-noise ratio (SNR). Let $N_t$ and $N_r$  denote the number of transmit and receive antennas, respectively.  Then the $N_r\times 1$ complex unitary vector $\bw_n$ represents the weights for maximum ratio combining,  and the $N_t\times 1$ complex unitary vector $\bff_n$ denotes the transmit beamforming vector. The $N_r\times N_t$ matrix $\bH_n$ represents the $n$th realization of the MIMO channel. The random process $\{\bH_n\}$ is assumed to be stationary and temporally correlated. Note that the common assumption of a complex Gaussian channel is unnecessary for our analysis.

The receiver continuously estimates the CSI sequence $\{\bH_n\}$ using pilot symbols sent by the transmitter. The estimated CSI is used for computing the maximum-ratio combining vector and the beamforming vector for feedback. This paper considers the scenario where CSI quantization and feedback delay are the main sources of transmit CSI inaccuracy. Thus, CSI estimation is assumed perfect for simplicity. This assumption is commonly made in the literature of limited feedback (see e.g. \cite{LovHeaETAL:GrasBeamMultMult:Oct:03,MukSabETAL:BeamFiniRateFeed:Oct:03}). Based on this assumption, the maximum-ratio combining vector is computed as $\bw_n = \bH_n\bff_n/\|\bH_n\bff_n\|$. In the next section, the beamforming vector $\bff_n$ is selected from a codebook for maximizing the receive SNR.

\subsection{CSI Feedback Link}\label{Section:System:Fb}
In Fig.~\ref{Fig:System}, CSI feedback link refers to the CSI path from the input of CSI quantizer to the output of the feedback channel. To satisfy a finite-rate feedback constraint, CSI is quantized efficiently by using a Grassmannian codebook, denoted as $\mathcal{F}$,  designed for beamforming in \cite{LovHeaETAL:GrasBeamMultMult:Oct:03, MukSabETAL:BeamFiniRateFeed:Oct:03}. Let $\bv_\ell$ denote the $\ell$th of $N$ unitary vectors in $\mathcal{F}$. To maximize the receive SNR, the quantizer function, denoted as $\mathcal{Q}_f$ (or $\mathcal{Q}_i$),  maps the channel matrix $\bH_n$ to a beamforming vector in the codebook $\mathcal{F}$ (or its index denoted as $I_n$) as follows \cite{LovHeaETAL:GrasBeamMultMult:Oct:03, MukSabETAL:BeamFiniRateFeed:Oct:03}
\begin{equation}\label{Eq:Quantizer}
\bff_n = \mathcal{Q}_f(\bH_n) = \arg\max_{\bv\in\mathcal{F}}\left\|\bH_n\bv\right\|^2, \quad I_n = \mathcal{Q}_i(\bH_n) = \arg\max_{1\leq \ell \leq N}\left\|\bH_n\bv_\ell\right\|^2.
\end{equation}
The feedback of the index $I_n$, called the \emph{channel state} hereafter, is sufficient for the transmitter to retrieve the selected beamforming vector $\bff_n$ from the codebook $\mathcal{F}$.

From \eqref{Eq:Quantizer}, the channel states, $\{I_n\}$,  are a sequence of alphabets of $N$ letters. This alphabet sequence is encoded by using a $B$-bit fixed-length code, where $B=\log_2N$ \cite{GallagerBook:InfoTheoReliableComm:68}. The CSI bit rate at the output of the source encoder is derived in Section~\ref{Section:CSI:Rate}. Intuitively, a more efficient alternative is to encode a long block of channel states by using variable-length codes such as a Huffman code \cite{GallagerBook:InfoTheoReliableComm:68}. Nevertheless, block CSI encoding contributes additional feedback delay, which decreases throughput significantly as shown in Section~\ref{Section:FbDelay}. Furthermore, a variable CSI codeword length is unsuitable for typical limited feedback systems such as 3GPP-LTE \cite{3GPP-LTE}, where an uniform number of bits are allocated for each instant of periodic CSI feedback. This also motivates the use of a periodic feedback protocol in this paper.

The periodic feedback protocol transmits the latest encoded CSI  to the feedback channel at fixed intervals. Thereby this protocol introduces variable CSI feedback delay as elaborated shortly. Let the integers $K$ and $D$ denote respectively a feedback interval and fixed feedback delay in samples, where $D$ is contributed by sources including signal processing and propagation. Then CSI used for beamforming in successive symbol durations lags behind the corresponding channel states by $D, (D+1), \cdots, (D+K-1)$ samples repeated in a cyclic order. The  CSI feedback bit rate for the periodic feedback protocol is derived under a feedback outage constraint in Section~\ref{Section:FbBandwidth}, and the effects of CSI feedback delay are investigated in Section~\ref{Section:FbDelay}.

To simplify analysis, the feedback channel is assumed free of errors, which is  typical in the literature of limited feedback (see e.g. \cite{LoveHeathBook, LovHeaETAL:GrasBeamMultMult:Oct:03, MukSabETAL:BeamFiniRateFeed:Oct:03}). This assumption is justified by the fact that beamforming feedback as  a control signal is usually  well protected by using error-correction coding or high  transmission power.

\section{Channel State Markov Chain }\label{Section:ChanMarkov}
In this section, the channel state sequence is modeled as a first-order finite-state Markov chain.  The channel state Markov chain is used as the primary analytical tool in the sequel. As discussed in Section~\ref{Section:Introd}, finite-state Markov chain models for fading have been established as valid models of wireless channels \cite{Wang:FinStatMarkovChan:95, Pimentel:FiniteMarkovModelCorrRicianFading:2004, Tan:MarkovRayleighFading:2000, Zhang:MarkovModelRayleighFading:1999, Babich:GenMarkovModelFlatFading:2000, Babich1999, Saadani:MarkovRayleighRake:2004, Guan:FSMCCorrFading:1999, Wang:VerifyFirstOrderMarkovRayleigh:1996, Sadeghi:CapAnalysisMarkovFlatFading:2005, Bischl:PacketErrorNonInterleaveRayleigh:1995, Hueda:InfoTheoTestMarkovBlockErrors:2005}, and widely used in wireless communication and networking \cite{Galluccio2005, Turin:MAPDecodErrorBurst:2001, Razavilar:OptimRateDelayControlFading:2002, Buche:ControlMobileCommTimeVarchan:2002, Yun:MarkovErrARQFading:2005, LiGoldsmith:MLDetectMarkov:2002}. Following this common approach, a finite-state Markov chain is used for modeling partial CSI of a MIMO fading  channel in this section. The resultant Markov chain is called the \emph{channel state Markov chain}. Its parameters, namely the state space and the stationary and transition probabilities, are related to the channel statistics and the CSI quantizer codebook. The channel state Markov chain model is validated using simulation in Section~\ref{Section:Numerical:VerifyMarkov}.

The channel state Markov chain models the time-variation of the channel state $I_n$ in \eqref{Eq:Quantizer}. Mapped from a stationary channel by the quantizer function in \eqref{Eq:Quantizer}, $\{I_n\}$ is a finite-sate stationary stochastic process. Motivated by the common approach in fading channel modeling \cite{Wang:FinStatMarkovChan:95, Pimentel:FiniteMarkovModelCorrRicianFading:2004, Tan:MarkovRayleighFading:2000, Zhang:MarkovModelRayleighFading:1999, Babich:GenMarkovModelFlatFading:2000},  this process is modeled as a homogeneous finite-state Markov chain of order one with the state space $\mathcal{I}=\{1,2, \cdots, N\}$ \cite{GallagerBook:StochasticProcs:95}. The key property of this Markov model is that conditioned on the most recent state $I_{n-1}$, $I_n$ is independent of the past states $I_{n-2}, I_{n-3}, \cdots$, thus
\begin{equation}\label{Eq:MarkovProp}
\Pr(I_n = \ell_n\mid I_{n-1} = \ell_{n-1}, I_{n-2} = \ell_{n-2},\cdots) = \Pr(I_n = \ell_n\mid I_{n-1} = \ell_{n-1}).
\end{equation}
Give the above property, the probability of a transition from state $m$ to $\ell$ is defined as $P_{\ell m}= \Pr(I_n = \ell \mid I_{n-1} = m )$. Moreover, the stationary probability of state $\ell$ is defined  as $\pi_{\ell} = \Pr(I_n = \ell)$. For convenience, denote the $N\times 1$ stationary probability vector as $\bpi$ with $[\bpi]_{\ell} = \pi_{\ell}$, and the $N\times N$ transition probability matrix as $\bP$ with $[\bP]_{\ell m} = P_{\ell m}$,  which is known as the \emph{stochastic matrix} \cite{GallagerBook:StochasticProcs:95}.


The channel state Markov chain is assumed to be \emph{ergodic} \cite{GallagerBook:StochasticProcs:95}. The ergodicity  implies three properties. First, the states of the  Markov chain are \emph{communicating}, namely that a transition between every pair of states occurs within a finite duration. Second, each state is \emph{recurrent} and thus the probability of returning to a same state is one. Third, each state is \emph{aperiodic}. This property exists if all possible time durations (in samples) of leaving and returning to each state have the common divisor of one. The assumption of ergodicity for the channel state Markov chain is valid for typical continuous channel distributions such as Rayleigh or Rician \cite{GoldsmithBook:WirelessComm:05}. This assumption provides the following property \cite{GallagerBook:StochasticProcs:95}
\begin{equation}\label{Eq:FSMC:Property}
\lim_{D\rightarrow\infty}\bP^D = \left[\bpi, \bpi, \cdots, \bpi\right].
\end{equation}

The probabilities $\pi_{\ell}$ and $P_{\ell m}$ depend on both the channel statistics and the CSI quantizer codebook. To specify these relationships, define a \emph{Voronoi cell } based on the quantizer in \eqref{Eq:Quantizer} as \cite{GerGra:VectQuanSignComp:92}
\begin{equation}
\mathcal{V}_{\ell} = \left\{\bH \in \mathds{C}^{N_r\times N_t}\mid \|\bH\bv_{\ell}\| \geq \|\bH\bv_m\|\ \forall \ m\in\mathcal{I}, m\neq \ell \right\}
\end{equation}
where $\bv_{\ell}$ denotes the $\ell$th unitary vector in the codebook $\mathcal{F}$. This set $\mathcal{V}_{\ell}$ maps the channel to the $\ell$th state of the channel state Markov chain as follows
\begin{equation}
I_n =\ell \Longleftrightarrow \bH_n \in\mathcal{V}_{\ell}.
\end{equation}
Using the above relationship, the probabilities of the channel state Markov chain can be related to the channel statistics as
\begin{equation}\label{Eq:MarkovPara}
\pi_{\ell} = \Pr\left(\bH \in \mathcal{V}_{\ell}\right), \quad P_{\ell m} =
    \Pr\left(\bH_n \in \mathcal{V}_{\ell}\mid \bH_{n-1} \in \mathcal{V}_m\right).
\end{equation}
In general, \eqref{Eq:MarkovPara} does not yield closed-form expression for the Markov chain probabilities except for the degenerate case of single antennas. Nevertheless, these equations are useful for computing the probabilities by simulation as discussed in the next section.

In this paper, the channels and hence the channel state Markov chains are assumed stationary. It is assumed that the receiver perfectly estimates the Markov-chain parameters, namely the stationary and transition probabilities. To facilitate CSI feedback, the receiver communicates once to the transmitter functions of the Markov chain parameters including the CSI feedback bit rate, the allowable feedback delay, and the source codes for compressing feedback CSI, which are derived in the sequel. Given channels that are stationary or approximately so, the infrequent feedback of these functions incurs overhead  that is much smaller than that for the instantaneous CSI feedback for beamforming, and hence is neglected in our analysis.

\section{CSI Source and Feedback Bit Rates}\label{Section:CSIRate:Bandwidth}
In this section, the overhead required for CSI feedback is analyzed based on the channel state Markov chain. Specifically, the CSI source and feedback bit rates are derived in  Section~\ref{Section:CSI:Rate} and Section~\ref{Section:FbBandwidth}, respectively.

\subsection{CSI Source Bit Rate}\label{Section:CSI:Rate}
The CSI source bit rate, defined as the average rate of encoded CSI bits, measures the amount of information generated by the temporally correlated MIMO channel. Based on fixed-length source coding as discussed in Section~\ref{Section:System}, the CSI source rate is the product of the CSI codeword length and the average rate of channel-state transition given as
\begin{equation}\label{Eq:AvRate:Time}
R_s = \lim_{M\rightarrow\infty}\frac{B}{MT}\sum_{n=1}^M 1\{I_n \neq I_{n-1}\}.
\end{equation}
Given the ergodicity  assumption for  the channel state Markov chain, $R_s$ in \eqref{Eq:AvRate:Time} can be rewritten in terms of the Markov chain probabilities by applying the ergodic theorem in Lemma~\ref{Lem:ErgTheo} from \cite[Corolary 4.1]{BremaudBook}.
\begin{lemma}\label{Lem:ErgTheo}\emph{Given a function
$z:\mathcal{I}\times\mathcal{I}\rightarrow \mathds{R_s}$ that satisfies
$\sum_{\ell=1}^{N}\sum_{m=1}^{N}|z(\ell,m)|\pi_{\ell}P_{\ell m} < \infty$,
the following convergence exists almost surely}
\begin{equation}
\lim_{M\rightarrow\infty}\frac{1}{M}\sum_{n=1}^{M}z(I_n,
I_{n-1})=\sum_{\ell=1}^{N}\sum_{m=1}^{N}z(\ell,m)\pi_{\ell}P_{\ell m}.
\end{equation}
\end{lemma}
The main result of this section as summarized in Proposition~\ref{Prop:AvRate} follows from \eqref{Eq:AvRate:Time} and Lemma~\ref{Lem:ErgTheo}. Note that the condition for applying Lemma~\ref{Lem:ErgTheo} is checked to be satisfied by setting $z(\ell, m) = 1\{\ell\neq m)$.
\begin{proposition}\label{Prop:AvRate}\emph{
The  CSI source bit rate is
\begin{equation}\label{Eq:AvRate:Erg}
R_s = \frac{B}{T}\sum_{\ell=1}^{N}\pi_{\ell}(1-P_{\ell\ell}).
\end{equation}
}
\end{proposition}
The above result shows that $R_s$ decreases linearly with increasing  probabilities $\{P_{\ell\ell}\}$, which characterize the degree of channel temporal correlation. By definition, $P_{\ell\ell}$  is the probability that the channel state  remains as  $\ell$ for two consecutive samples. The values of $\{P_{\ell\ell}\}$ are close to one  if the channel is highly correlated in time, or close to zero for fast fading. As shown by simulation in Section~\ref{Section:Numerical:FbRate}, $R_s$ increases linearly with Doppler shift for spatially i.i.d. Rayleigh fading and the Clarke's fading model \cite{GoldsmithBook:WirelessComm:05}.

\subsection{CSI Feedback Bit Rate}\label{Section:FbBandwidth}
The \emph{CSI feedback bit rate} is defined as the maximum bit rate supported by the feedback channel. In this section, the CSI feedback bit rate is derived under a constraint on feedback outage probability. Recall that a feedback outage refers to the event that more than one channel-state transition occurs within one feedback interval. Let $Z$ denote the number of channel-state transitions in a feedback interval of $K$ samples. Moreover, let $P_o$ represent the feedback outage probability, thus $P_o := \Pr(Z > 1 )$. For a small real number $\delta > 0$, the constraint $P_o \leq \delta$ ensures that CSI feedback is sufficiently frequent.

The main result of this section is obtained and summarized in the in following proposition.
\begin{proposition}\label{Prop:FbBW}\emph{
Under the feedback outage constraint $P_o \leq \delta$, the CSI feedback bit rate is $R_f = \frac{B}{KT}$ bit/s  where the feedback interval in samples, $K$,  is the largest integer subject to
\begin{equation}\label{Eq:FbConstraint:a}
\sum_{\ell=1}^NP^K_{\ell\ell}\pi_{\ell} + \sum_{\ell=1}^N\sum_{\substack{m=1\\ m\neq \ell}}^N\frac{(P_{mm}^K - P_{\ell\ell}^K)P_{m\ell}\pi_{\ell}}{P_{mm}-P_{\ell\ell}}\geq 1 - \delta.
\end{equation}
}
\end{proposition}
\begin{proof}
See Appendix~\ref{App:FbBW}.
\end{proof}
To observe the dependence of the CSI feedback bit rate on channel temporal correlation, bounds on $R_f$ can be obtained as a by-product of the proof for Corollary~\ref{Cor:FbBw}
\begin{equation}
\frac{\log(\max_{a}P_{aa})}{\log(1-\delta) - \log\left(1 + \sum_{m\neq \ell}\frac{P_{m\ell}\pi_{\ell}}{|P_{mm}-P_{\ell\ell}|}\right)}\leq R_f \leq \frac{\log(\min_a P_{a a})}{\log(1-\delta)}.
\end{equation}
The numerators of the above bounds suggest that $R_f$ decreases with increasing probabilities $\{P_{\ell\ell}\}$, which characterize the degree of channel temporal correlation as mentioned earlier. In other words, $R_f$ is smaller for a more temporally correlated channel and vise versa, which is also observed  for the CSI source bit rate given in Proposition~\ref{Prop:AvRate}.

The feedback interval $K$ for the extreme cases of small and large target feedback outage  probabilities are characterized in the following corollary of Proposition~\ref{Prop:FbBW}.
\begin{corollary}\label{Cor:FbBw}\emph{For a small target feedback outage probability $\delta \rightarrow 0$, the normalized feedback interval converges to one: $K\rightarrow 1$; for a large probability $\delta \rightarrow 1$, $K$ scales as
\begin{equation}\label{Eq:K:Extreme}
    K = \eta \log(1-\delta) + O(1)
\end{equation}
where $\frac{1}{\log\max_{\ell}P_{\ell\ell}}\leq\eta \leq \frac{1}{\log\min_{\ell}P_{\ell\ell}}$.
}
\end{corollary}
\begin{proof}
See Appendix~\ref{App:FbBw:Cor}.
\end{proof}
For $\delta\rightarrow 1$, the first term $\eta \log(1-\delta) $ in \eqref{Eq:K:Extreme} is dominant because it is asymptotically large. The factor $\eta$ in \eqref{Eq:K:Extreme} represents the sensitivity of $K$ to the change on the feedback constraint $\delta$. This factor depends on the degree of channel temporal correlation through the probabilities $\{P_{\ell\ell}\}$.

\section{Feedback Delay}\label{Section:FbDelay}
This section focuses on the effects of feedback delay on the throughput of the limited feedback beamforming system. We consider both \emph{fixed delay} due to signal processing and propagation, and \emph{variable delay} caused by the periodic feedback protocol. In Section~\ref{Section:FbDelay:Cap}, the feedback throughput gain with fixed feedback delay is defined and derived. In Section~\ref{Section:FbDelay:Fixed}, fixed feedback delay is shown to reduce the feedback throughput gain at least at an exponential rate. Building on this result, an upper-bound on the feedback throughput gain is derived as a function of both fixed and variable feedback delay in Section~\ref{Section:FbDelay:Var}.


\subsection{Feedback Throughput Gain}\label{Section:FbDelay:Cap}
The feedback throughput gain is defined as the gain in ergodic throughput due to delayed CSI feedback with respect to the case of infinite feedback delay. In this section, only fixed feedback delay is considered and a corresponding upper bound on the feedback throughput gain is derived. This result is extended in Section~\ref{Section:FbDelay:Var} to include the effect of variable feedback delay. Let $D$ denote fixed feedback delay in samples, and $R$ represent ergodic throughput. Then the feedback throughput gain  can be written as  $\Delta R(D) = R(D) - R(\infty)$.

The ergodic throughput $R(D)$ is derived as follows. With feedback delay, the input-output relationship for the limited feedback beamforming system in \eqref{Eq:Sys} can be rewritten as
\begin{equation}\label{Eq:SISO}
y_n = \sqrt{P}g_n(D) x_n + \nu_n
\end{equation}
where the effective channel gain $g_n(D) = \|\bH_n \bff(I_{n-D})\|$ and $\bff(I_{n-D})$ is the transmit beamforming vector chosen based on the delayed feedback CSI $I_{n-D}$. With only beamforming feedback, $g_n(D)$ is unknown to the transmitter, and hence constant transmission power $P$ is optimal \cite{GoldsmithBook:WirelessComm:05}. Given feedback delay, deriving the optimal strategy for transmit beamforming  is difficult as it depends on the codebook, the channel stochastic distribution, and the channel prediction algorithm. For simplicity, this paper adopts the strategy of applying the codebook vector corresponding to $I_{n-D}$ as the beamforming vector. Equivalently,  $\bff(I_{n-D}) = \mathcal{Q}_f(\bH_{n-D})$ with the quantizer function $Q(\cdot)$ given in \eqref{Eq:Quantizer}.  Note that this sub-optimal beamforming strategy is optimal for zero feedback delay \cite{LovHeaETAL:GrasBeamMultMult:Oct:03}. Using this strategy, the ergodic throughput of the effective  scalar channel in  \eqref{Eq:SISO} is obtained as \cite{GoldsmithBook:WirelessComm:05}
\begin{equation}\label{Eq:ErgCap}
R(D) = \E\left[\log_2(1+g_n^2(D))\right] = \E\left[\log_2(1+P\|\bH_n \mathcal{Q}_f(\bH_{n-D})\|^2)\right].
\end{equation}
To achieve this throughput, the channel code for forward-link data is assumed to be sufficiently long to attain channel ergodicity \cite{GoldsmithBook:WirelessComm:05}. In other words, the code length and hence the decoding delay are much longer than channel coherence time. If a constraint on the decoding delay is required, the outage capacity is a more appropriate performance metric \cite{Ozarow:InfoTheoCellularMobile:1994}.

The ergodic throughput in \eqref{Eq:ErgCap} can be rewritten in terms of the probabilities of the channel state Markov chain as shown in the following lemma.
\begin{lemma}\label{Lem:ErgCap}\emph{For fixed feedback delay of $D$ symbol durations, the ergodic throughput is \begin{equation}
\label{Eq:ErgCap:a}
    R(D) = \sum_{\ell=1}^N\sum_{m=1}^NR_{\ell m}\left[\bP^D\right]_{\ell m}\pi_m
\end{equation}
where
\begin{equation}\label{Eq:ErgCap:Comp}
R_{\ell m}= \E\left[\log_2(1+P\|\bH_n \bv_m\|^2)\mid \bH_n\in\mathcal{V}_{\ell}, \bH_{n-d}\in\mathcal{V}_m\right].
\end{equation}
}
\end{lemma}
\begin{proof}
See Appendix~\ref{App:ErgCap}.
\end{proof}
In general, the constants $\{R_{\ell m}\}$ do not have closed-form expressions and have to be estimated by using Monte Carlo simulation.

Consider the extreme cases of zero delay ($D=0$) and infinite delay ($D\rightarrow \infty$). The corresponding ergodic throughput is obtained and shown in the following corollary of Lemma~\ref{Lem:ErgCap}.
\begin{corollary}\label{Cor:ErgCap:SpecCases}\emph{
The ergodic throughput for zero feedback delay $R(0)$ and infinite delay $R(\infty)$ are }
\begin{eqnarray}
    R(0) &=& \sum_{\ell=1}^N R_{\ell \ell}\pi_{\ell} = \E\left[\log_2\left(1+P\|\bH\mathcal{Q}_f(\bH)\|^2\right)\right]   \label{Eq:ErgCap:NoDly}\\
    R(\infty) &=& \sum_{\ell=1}^N\sum_{m=1}^NR_{\ell m}\pi_{\ell}\pi_m =\sum_{\ell=1}^N \E\left[\log_2\left(1+P\|\bH\bv_{\ell}\|^2\right)\right]\pi_{\ell}. \label{Eq:ErgCap:InfDly}
\end{eqnarray}
\end{corollary}
\begin{proof}
See Appendix~\ref{App:ErgCap:SpecCases}.
\end{proof}
A few remarks are in order.
\begin{enumerate}
\item For zero feedback delay, $R(0)$ is equal to the ergodic capacity for limited feedback beamforming based on the block fading channel model \cite[(26)]{LovHeaETAL:GrasBeamMultMult:Oct:03}, where feedback delay is omitted.

\item For infinite feedback delay, the beamforming vector at the transmitter is adapted to obsolete CSI independent of the current channel state. Higher ergodic throughput than $R(\infty)$  may be achieved by using space time block coding or adapting the beamforming vector to channel statistics \cite{PaulrajBook}, both require no instantaneous CSI. Despite its sub-optimality, $R(\infty)$ serves as a reference value for computing the feedback throughput gain in this paper.

\item It can be observed from \eqref{Eq:ErgCap:NoDly} and \eqref{Eq:ErgCap:InfDly} that both $R(0)$ and $R(\infty)$ are independent of the transition probabilities of the channel state Markov channel. This suggests that channel dynamics affect only the case of finite feedback delay as considered in the next section.
\end{enumerate}

Finally, from Lemma~\ref{Cor:ErgCap:SpecCases}, Corollary~\ref{Cor:ErgCap:SpecCases} and the definition, the feedback throughput gain is readily written as shown in the following proposition.
\begin{proposition}\label{Prop:FbCapGain:Def}\emph{Assuming only fixed feedback delay, the feedback throughput gain is given as
\begin{equation}\label{Eq:FbCapGain:Def}
\Delta R(D) = \sum_{\ell=1}^N\sum_{m=1}^NR_{\ell m}\left\{\left[\bP^D\right]_{\ell m} - \pi_l\right\}\pi_m
\end{equation}
where $R_{\ell m}$ is given in \eqref{Eq:ErgCap:Comp}.
}
\end{proposition}
An upper-bound on $R(D)$ is derived in the next section.

{\bf Other metrics:} Besides the feedback throughput gain, other metrics can be defined for quantifying the effects of CSI quantization and feedback delay on the throughput of a transmit beamforming system. The ergodic capacity for perfect CSI feedback is $C_{\text{ideal}} = \E\left[\log_2\left(1+P\lambda^2_1(\bH)\right)\right]$ where $\lambda_1(\bH)$ represents the largest singular value of $\bH$. With respect to the case of perfect CSI feedback, the capacity loss due to CSI quantization, called \emph{quantization loss}, can be defined as $\nabla\!\!_q C = C_{\text{ideal}} - R(0)$ where $R(0)$ follows from \eqref{Eq:ErgCap}. As shown in \cite{LovHeaETAL:GrasBeamMultMult:Oct:03, MukSabETAL:BeamFiniRateFeed:Oct:03}, quantization loss is small given just a few bits of resolution for quantized CSI. Moreover, the capacity loss due to both CSI quantization and feedback delay can be written as $\nabla\!\!_{qd} R = C_{\text{ideal}} - R(D)$. Next, using the case of infinite feedback delay as the reference, the \emph{maximum throughput gain} is defined as $\Delta_{\max}R = C_{\text{ideal}} - R(\infty)$. The feedback throughput gain defined earlier as $\Delta R(D) = R(D) - R(\infty)$ takes into account of both CSI quantization and feedback delay. By definition, the feedback throughput gain $\Delta R(D)$ is the complement of the capacity loss $\nabla\!\!_{qd} R$ in the sense that $\Delta R(D) + \nabla\!\!_{qd} R = \Delta_{\max}C$.

\subsection{Effect of Fixed Feedback Delay}\label{Section:FbDelay:Fixed}
In this section, increasing fixed feedback delay is shown to reduce the feedback throughput gain at least at an exponential rate. This  result is derived based on theory of Markov chain convergence rate. To state this theory, define the \emph{time reversal} of the stochastic matrix $\bP$, denoted by $\tilde{\bP}$, as a  matrix whose components are given as \cite{GallagerBook:StochasticProcs:95}
\begin{equation}\label{Eq:TranMat:Reverse}
\tilde{P}_{m\ell } = \frac{\pi_m}{\pi_{\ell}}P_{\ell m}, \quad  1 \leq m, \ell \leq N.
\end{equation}
As implied by its name, $\tilde{\bP}$ represents channel-state transitions in the reverse direction of $\bP$. For the special case of  $\tilde{\bP}=\bP$, the channel state Markov chain is known as \emph{reversible} \cite{GallagerBook:StochasticProcs:95}. Using the definition in \eqref{Eq:TranMat:Reverse}, \cite[Theorem~2.1]{Fill91:EigenvalBndsNonrevMarkovChain:91} is restated as the following lemma.
\begin{lemma}\label{Lem:RedRate}\emph{
For the ergodic channel state Markov chain, the following inequality holds
\begin{equation}\label{Eq:PrVarDist}
    \left(\sum_{\ell=1}^N\left|\left[\bP^D\right]_{\ell m}-\pi_{\ell}\right|\right)^2 \leq \frac{\lambda^D}{\pi_m}, \quad 1\leq m \leq N
\end{equation}
where $\lambda\in[0,1]$ is the second largest eigenvalue of the matrix $\bP\tilde{\bP}$.
}
\end{lemma}

By using Lemma~\ref{Lem:RedRate}, the main result of this section is derived and shown in the following proposition.
\begin{proposition}\label{Prop:FbCapGain}\emph{
For feedback delay of $D$ samples, the feedback throughput gain is upper bounded as
\begin{equation}\label{Eq:GainRate}
    0\leq \Delta R(D) \leq \alpha(\sqrt{\lambda})^D
\end{equation}
where $\alpha = \sum_{m=1}^N\sqrt{\pi_m}\max_{\ell}R_{\ell m}$ with $R_{\ell m}$ given in \eqref{Eq:ErgCap:Comp}, and $\lambda\in[0,1]$ is the second largest eigenvalue of the matrix $\bP\tilde{\bP}$.}
\end{proposition}
\begin{proof}
See Appendix~\ref{Appendix:GainRate}.
\end{proof}
A few further remarks are in order.
\begin{enumerate}
\item The eigenvalue $\lambda$ is a key parameter characterizing the channel dynamics. A larger value of $\lambda$ indicates longer channel coherence time and vise versa.

\item As observed from \eqref{Eq:GainRate}, the feedback throughput gain decreases at least exponentially with the feedback delay. The decreasing rate is $\sqrt{\lambda}$ and thus depends on channel coherence time.

\item For a reversible channel state Markov chain with $\bP=\tilde{\bP}$,  $\sqrt{\lambda}$ is the second largest eigenvalue of the transition matrix    $\bP$.

\item As observed from simulation results in Section~\ref{Section:Numerical:FbDelay}, the upper bound in \eqref{Eq:GainRate} is tight in several cases of interest.
\end{enumerate}

\subsection{Effects of both Fixed and Variable Feedback Delay}\label{Section:FbDelay:Var}
In this section, the effects of both fixed and variable feedback delay on the feedback throughput gain are jointly considered. As discussed in Section~\ref{Section:System:Fb}, due to the periodic feedback protocol, CSI used for beamforming in successive  symbol durations encounters variable feedback delay of $D, (D+1), \cdots, (D+K-1)$ samples repeated in a cyclic order. Let $\tilde{R}(K, D)$ denote the ergodic throughput with both fixed and variable feedback delay. Thus, $\tilde{R}(K, D)$  can be written as
\begin{equation}\label{Eq:ErgCap:PeriFb}
\tilde{R}(K, D) = \frac{1}{K}\sum_{k=0}^{K-1}R(k+D)
\end{equation}
where $R(\cdot)$  is given in \eqref{Eq:ErgCap:a}. Define the corresponding feedback throughput gain as $\Delta \tilde{R} = \tilde{R}(K, D) - R(\infty)$. Then we have the following corollary of Proposition~\ref{Prop:FbCapGain}.
\begin{corollary}\label{Cor:FbCapGain:Compr}\emph{
For fixed feedback delay of $D$  samples and a feedback interval of $K$ samples, the feedback throughput gain is upper bounded as
\begin{equation}\label{Eq:GainRate:VarDly}
    0\leq \Delta \tilde{R}(K, D) \leq \alpha(\sqrt{\lambda})^{D}\frac{1-(\sqrt{\lambda})^K}{K(1-\sqrt{\lambda})}.
\end{equation}
where $\alpha$ and $\lambda$ are identical to those in Proposition~\ref{Prop:FbCapGain}.}
\end{corollary}
The upper-bound in \eqref{Eq:GainRate:VarDly} can be decomposed into three factors $\alpha$, $(\sqrt{\lambda})^{D}$  and $\frac{1-(\sqrt{\lambda})^K}{K(1-\sqrt{\lambda})}$. They characterize the effects of CSI quantization, fixed feedback delay and the periodic feedback protocol, respectively. In particular, the feedback throughput gain decreases at least exponentially with $D$ and approximately inversely with $K$. Thus, for small values of $\lambda$ corresponding to fast fading, the fixed feedback delay has a more significant effect on throughput than the delay due to the feedback protocol.

Define the normalized feedback throughput gain as $\Delta \bar{R}(K, D) = \Delta \tilde{R}(K, D)/\Delta R(0)$, where the normalization factor $\Delta R(0)$ corresponds to the ideal case of zero feedback delay. Motivated by the result in Corollary~\ref{Cor:FbCapGain:Compr}, $\Delta \bar{R}(K, D)$ can be approximated as
\begin{equation}\label{Eq:FbCapGain:Approx}
\Delta \bar{R}(K, D) \approx \underbrace{(\sqrt{\lambda})^{D}}_{q_D}\times \underbrace{\frac{\left[1-(\sqrt{\lambda})^K\right]}{K(1-\sqrt{\lambda})}}_{q_K}
\end{equation}
where the factors $q_D$ and $q_K$  separate the effects of the fixed and variable feedback delay on the feedback throughput gain. For a sanity check, the above approximated expression gives 1 for zero feedback delay $K=D=0$, and zero for infinite delay $K\rightarrow \infty$ or $D\rightarrow \infty$. The approximation of the normalized feedback throughput gain in \eqref{Eq:FbCapGain:Approx} is observed to be accurate from simulation results in Section~\ref{Section:Numerical}. The result in \eqref{Eq:FbCapGain:Approx} is useful for computing the allowable feedback delay under a constraint on the normalized feedback throughput gain as demonstrated by the design example in Section~\ref{Section:Numerical:DesignEx}.

\section{Feedback Compression}\label{Section:FBCompress}
An algorithm is proposed in Section~\ref{Section:FBCompress:Algo} for compressing CSI feedback by exploiting temporal correlation in feedback CSI. The effect of feedback compression on feedback throughput gain is analyzed in Section~\ref{Section:FBCompress:Cap}.

\subsection{Algorithm}\label{Section:FBCompress:Algo}
The feedback compression algorithm is motivated by the existence of residual temporal-correlation in feedback CSI, which is largely reduced by the periodic feedback protocol (cf. Section~\ref{Section:FbBandwidth}). To characterize such residual redundancy, feedback CSI is modeled as a Markov chain obtained by down-sampling the channel stat Markov chain at feedback intervals. Consequently, the stochastic matrix for this down-sampled Markov chain is $\bP^K$ with the $(\ell, m)th$ component denoted as $P^{(K)}_{\ell m}$. The correlation between two consecutive instants of feedback CSI is reflected in an uneven distribution of the transition probabilities in $\bP^K$. Specifically, conditioned on the previous instant of feedback CSI, the current one belongs to a subset of the Markov chain state space with high probability. To define this subset, let the transition probabilities for the Markov chain state $m$, namely $\left\{P^{(K)}_{\ell m}\right\}_{\ell=1}^N$, be indexed according to the descending order of their values, thus $P^{(K)}_{1 m}\geq P^{(K)}_{2m}\cdots \geq P^{(K)}_{Nm}$. For a small positive real number $\epsilon$, the high-probability subset, called the \emph{$\epsilon$-neighborhood}, of  the $m$th Markov chain state is defined as
\begin{equation}\label{Eq:Neighborhood}
    \mathcal{N}_m(\epsilon) = \left\{\tilde{N}\leq \ell \leq N \left| \sum_{\ell=\tilde{N}}^N  P^{(K)}_{\ell m} \geq 1- \epsilon, \sum_{\ell=\tilde{N}+1}^N  P^{(K)}_{\ell m} <  1- \epsilon\right.\right\}.
\end{equation}
This $\epsilon$-neighborhood groups  most likely transitions from the Markov chain state $m$ with total probability larger than  $(1-\epsilon)$. High channel temporal correlation results in small $\epsilon$-neighborhoods and vise versa.

Next, the feedback compression algorithm compresses feedback CSI in alternate feedback intervals to prevent propagation of CSI errors due to lossy compression. The possibility of error propagation is due to backward dependence of CSI compression in time. Conditioned on the $\tilde{n}$th feedback channel state $I_{\tilde{n}K} = m$ that is uncompressed, the next one $I_{(\tilde{n}+1)K}$ is compressed by lossy source coding. Specifically, $I_{(\tilde{n}+1)K}$ is encoded into one of $|\mathcal{N}_m(\epsilon)|$ fixed-length codewords if $I_{(\tilde{n}+1)K}$ belongs to the $\epsilon$-neighborhood of $I_{\tilde{n}K}$, namely $\mathcal{N}_m(\epsilon)$ defined in \eqref{Eq:Neighborhood}. Otherwise, a codeword indicating CSI truncation is generated by the source encoder.  It follows that compressing $I_{(\tilde{n}+1)K}$ requires a fixed-length source codebook having a size of  $[|\mathcal{N}_m(\epsilon)|+1]$ and a codeword length of $B_m(\epsilon) = \lceil\log_2(|\mathcal{N}_m(\epsilon)|+1)\rceil$ bits. The above source coding algorithm compresses CSI since $B_m(\epsilon)$ is usually much smaller than $B=\log_2N$ for the case of no compression. The number of bits for compressed CSI feedback should be uniform and thus is chosen as $\tilde{B}(\epsilon) = \max_{\ell}B_{\ell}(\epsilon)$. Finally,  to decode both compressed and uncompressed CSI feedback, the transmitter stores all $(N+1)$ source codebooks including one for uncompressed and $N$ for compressed CSI feedback. On decoding the codeword indicating CSI truncation, transmitter randomly chooses a beamforming vector from the codebook $\mathcal{F}$.

A few remarks are in order.
\begin{enumerate}
\item The proposed CSI compression algorithm alternates $B$ and $\tilde{B}(\epsilon)$ feedback bits for successive feedback instants, and hence the CSI compression ratio is
    \begin{equation}\label{Eq:CompRatio}
        \Delta R_f = \frac{B-\tilde{B}(\epsilon)}{2B}.
    \end{equation}
    This ratio is evaluated by simulation in Section~\ref{Section:Numerical:FbCompress}.

\item The proposed algorithm can be also applied for compressing other types of feedback CSI such as channel-gain feedback useful for power control and scheduling. Different types of compressed feedback CSI can be integrated such that the total number of CSI bits per feedback instant remains constant, which suits practical systems such as 3GPP-LTE \cite{3GPP-LTE}.

\item As mentioned earlier, the proposed feedback compression algorithm is preferred to the conventional lossless data compression algorithms, which use variable-length source coding and optionally block processing for higher compression efficiency \cite{GallagerBook:InfoTheoReliableComm:68, SimonLeus:FbReductionSMLinearPrecod:2007}. The reason is that variable-length CSI feedback is unsuitable for practical systems. Furthermore,  block processing causes additional feedback delay that significantly decreases throughput as shown in Section~\ref{Section:FbDelay:Fixed}.

\item The proposed feedback compression algorithm complements the periodic feedback protocol on further decreasing the CSI feedback rate by exploiting channel temporal correlation. In contrast, based on an aperiodic feedback protocol, the original version of this algorithm presented in \cite{Huang:LimitedFeedbackCompression:Globecom06} is the only feedback compression function in the CSI feedback link.

\item In practice, the receiver estimates the parameters of the channel-state Markov chain and computes the $N$ source codebooks for compressing feedback CSI. These codebooks are sent to the transmitter for decoding the compressed CSI. For stationary channels, one-time codebook feedback is sufficient.

\end{enumerate}

\subsection{Feedback Throughput Gain with Feedback Compression}\label{Section:FBCompress:Cap}
The CSI feedback compression algorithm proposed in the preceding section is lossy and hence inevitably  incurs loss on the feedback throughput gain. To characterize this loss, the ergodic throughput with feedback compression,  denoted as $\hat{R}(K, D)$, is written as
\begin{equation}\label{Eq:ErgCap:Compr}
\hat{R}(K, D) = \frac{1}{2}R(K,D) + \frac{(1-\epsilon)}{2}R(K,D) + \frac{\epsilon }{2}R(\infty)
\end{equation}
where $R(K,D)$ is the ergodic throughput without feedback compression. The three terms on the right-hand-side of \eqref{Eq:ErgCap:Compr} correspond to uncompressed, compressed, and truncated feedback CSI, respectively. Note that the probability for CSI truncation is $\epsilon$. The feedback throughput gain with feedback compression is obtained by using \eqref{Eq:ErgCap:Compr} and given in the following proposition. Also shown is the CSI feedback bit rate for CSI compression.
\begin{proposition}\label{Prop:FbCompr}\emph{The proposed CSI feedback compression algorithm has the following properties. }
\begin{enumerate}
\item The feedback throughput gain is
\begin{equation}\label{Eq:FbCapGain:Compr}
\Delta \hat{R}(K, D, \epsilon) = \left(1-\frac{\epsilon}{2}\right)\Delta R(K, D)
\end{equation}
where $\Delta R(K, D)$ corresponds to the case of no feedback compression.

\item The average CSI feedback bit rate is
\begin{equation}\label{Eq:FbRate:Compr}
\hat{R}_f(\epsilon) = \frac{B + \hat{B}(\epsilon)}{2KT}.
\end{equation}
\end{enumerate}
\end{proposition}
From \eqref{Eq:FbCapGain:Compr}, $\epsilon$ should be small so as to minimize the throughput loss due to lossy feedback CSI compression. Nevertheless, too small $\epsilon$ decreases compression efficiency since $\hat{R}_f(\epsilon)$ converges to the feedback bit rate of uncompressed CSI, $R_f$,  as $\epsilon$ reduces to zero. According to \eqref{Eq:FbRate:Compr}, $\hat{R}_f(\epsilon)$ is at least half of $R_f$.

\subsection{Extension: Block Feedback Compression}\label{Section:FBCompress:Ext}
For the feedback compression algorithm proposed in Section~\ref{Section:FBCompress:Algo}, compressing every other feedback CSI instant is usually sufficient since temporal correlation in CSI is substantially reduced by the periodic feedback protocol. Nevertheless, if the feedback interval is small, the residual correlation is significant and can be fully exploited by compressing CSI for multiple consecutive feedback instants, called \emph{block feedback compression}. The algorithm in Section~\ref{Section:FBCompress:Algo} can be extended to block feedback compression as follows. Let $W$ denote the compression block length in feedback interval $KT$. Feedback compression specified in Section~\ref{Section:FBCompress:Algo} repeats for a block of $W$ feedback instants. Upon the first occurrence of feedback truncation in compressing a CSI block,  the codewords indicating truncation are sent for the remaining feedback instants in this block. Uncompressed CSI feedback is performed between every two blocks of compressed CSI feedback to stop propagation of CSI errors due to feedback truncation.

For block feedback compression, the ergodic throughput in \eqref{Eq:ErgCap:Compr} is rewritten as
\begin{equation}\label{Eq:ErgCap:BlkCompr}
\hat{R}(K, D) = \frac{1}{W+1}R(K,D) + \frac{1}{W+1}\sum_{\ell =1}^W\left\{(1-\epsilon)^\ell R(K,D) + \left[1-(1-\epsilon)^\ell\right] R(\infty)\right\}.
\end{equation}
Using \eqref{Eq:ErgCap:BlkCompr}, the feedback throughput gain is obtained as shown in the following lemma.
\begin{lemma}\label{Prop:BlkFbCompr}\emph{For block feedback compression, the feedback throughput gain is}
\begin{equation}\label{Eq:FbCapGain:BlkCompr}
\Delta \hat{R}(K, D, \epsilon) = \frac{1-(1-\epsilon)^{W+1}}{(W+1)\epsilon}\Delta R(K, D)
\end{equation}
and the average CSI feedback bit rate is
\begin{equation}\label{Eq:FbRate:BlkCompr}
\hat{R}_f(\epsilon) = \frac{B + E\hat{B}(\epsilon)}{(E+1)KT}.
\end{equation}
\end{lemma}
As observed from \eqref{Eq:FbCapGain:BlkCompr}, if the temporal correlation in feedback CSI is high ($\epsilon << 1$), $\Delta \hat{R}(K, D, \epsilon) \approx \Delta R(K, D)$ and hence block feedback compression has a negligible effect on feedback throughput gain. Otherwise, the block length  should be set as $W=1$ to avoid throughput loss. According to \eqref{Eq:FbRate:BlkCompr}, the block feedback compression ratio can be up to $1/(E+1)$ instead of $50\%$ for the original algorithm in Section~\ref{Section:FBCompress:Algo}.

\section{Numerical Results and Design Example} \label{Section:Numerical}
In this section, numerical results and a design example are provided to obtain further insights into designing limited feedback beamforming systems.
The channel state Markov chain  is constructed based on the spatially i.i.d.  Rayleigh fading $4\times 4$ MIMO channel model, corresponding to rich scattering and the size of antenna array equal to four. In this model, each channel coefficient is $\mathcal{CN}(0,1)$. The temporal correlation of each channel coefficient is assumed to follow Clarke's model \cite{GoldsmithBook:WirelessComm:05} and is characterized by the zeroth order Bessel function of the first kind  $\mathcal{J}_0(2\pi f_D\tau)$, where $f_D$ is Doppler shift and $\tau$ is the time difference between two samples of a channel coefficient \cite{GoldsmithBook:WirelessComm:05}. Based on the above channel model, a first-order channel state Markov chain is constructed by simulation.

\subsection{Accuracy of Channel State Markov Chain}\label{Section:Numerical:VerifyMarkov}

In this section, the autocorrelation of quantized CSI is used as the criterion for validating the first-order channel state Markov chain. It is worth mentioning that the accuracy of this model is also supported by measurement data \cite{Daniels07:QuantizedFbReductionSMLinearPrecod:2007}, where the feedback throughput gain due to transmit beamforming has been measured by using a MIMO-OFDM prototype over an indoor wireless channel. The measurement data closely matches the theoretical result in \eqref{Eq:FbCapGain:Approx} derived using the channel state Markov chain.

\begin{figure}
\centering
  \includegraphics[width=9.5cm]{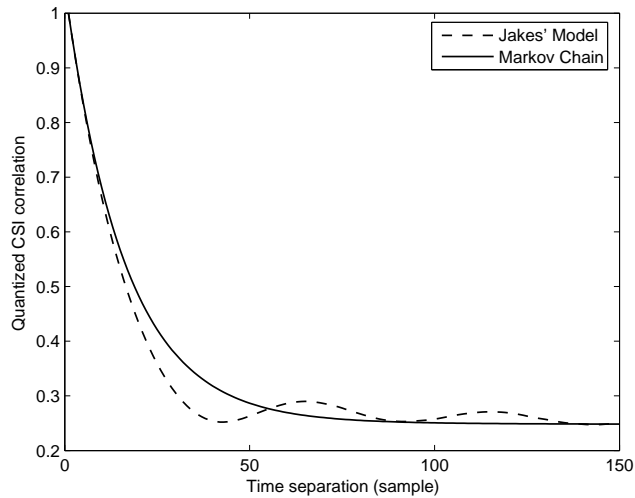}
  \caption{CSI autocorrelation versus time separation for the normalized Doppler shift $f_DT=10^{-2}$. For the ``Clarke's model", the autocorrelation function of each channel coefficient is given by the Clarke's function; for the ``Markov chain", the autocorrelation of quantized CSI is modeled by a first-order Markov chain. }\label{Fig:Crr}
\end{figure}

The quantized CSI for transmit beamforming is a correlated sequence of unitary vectors $\{\bff_n\}$. From the quantizer functions in \eqref{Eq:Quantizer}, it is clear that the CSI sequence  $\bff_n$ is one-to-one mapped to the channel state sequence $\{I_n\}$. The autocorrelation function of the CSI sequence is defined as $A(\hat{\tau}) = \E\left[\left|\bff_{n}^* \bff_{n+\hat{\tau}}\right|^2\right]$ where $\hat{\tau}$ is time separation in sample. The autocorrelation function can be evaluated by 1) simulating the $4\times 4$ MIMO channel as described earlier, 2) generating the quantized CSI sequence by applying \eqref{Eq:Quantizer}, and 3) computing autocorrelation function based on channel ergodicity, namely $A(\hat{\tau}) = \lim\limits_{M\rightarrow\infty}\frac{1}{M}\sum_{n=1}^M\left|\bff_{n}^* \bff_{n+\hat{\tau}}\right|^2$. The autocorrelation function such evaluated is plotted in Fig.~\ref{Fig:Crr} with the legend ``Clarke's model". The normalized Doppler shift is $f_DT = 10^{-2}$. By proper scaling the time separation ($x$-axis), Fig.~\ref{Fig:Crr} is also applicable for other normalized Doppler shifts.

Based on the channel state Markov chain, the CSI autocorrelation function is given by  \[A(\hat{\tau}) = \sum_{\ell=1}^N\sum_{m=1}^N \left|\bv_{\ell}^* \bv_{m}\right|^2 \left[\bP^{\hat{\tau}}\right]_{\ell m}\pi_m\] where $\bv_{\ell}$ is the $\ell$th vector in the codebook $\mathcal{F}$. The above autocorrelation is plotted in  Fig.~\ref{Fig:Crr} and labeled by the legend ``Markov chain".

As observed from Fig.~\ref{Fig:Crr}, the first-order Markov chain provides an good approximation of the Clarke's fading model. In particular, the approximation is very accurate for correlation larger than $0.5$. Moreover, both curves in Fig.~\ref{Fig:Crr} converge as the time separation increases. Fig.~\ref{Fig:Crr} suggests that higher-order  channel-state Markov chains practically provide no improvement on modeling accuracy. For instance, two curves deviate significantly only for time separation larger than $10$ samples. Consequently, a Markov chain of an order larger than $10$ is required to provide an significantly better approximation of Clarke's model. Unfortunately, such a high-order Markov chain is impractical as its complexity increases exponentially with the order.


\subsection{CSI Source and Feedback Bit Rates}\label{Section:Numerical:FbRate}

\begin{figure}
\centering
  \includegraphics[width=9.5cm]{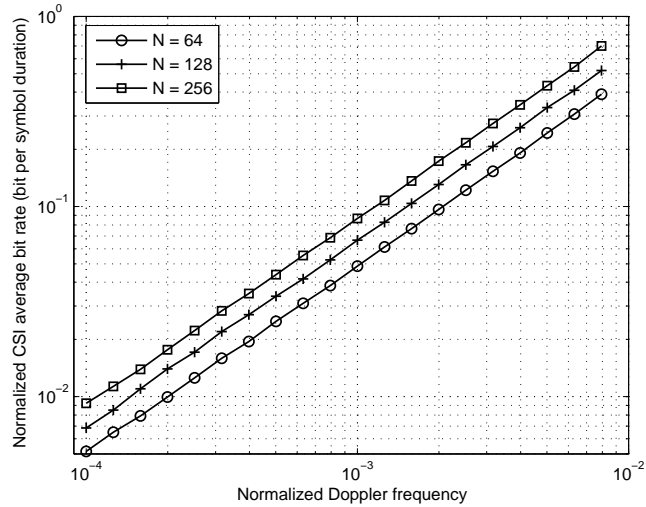}\\
  \caption{Normalized CSI source bit rate versus normalized Doppler shift for the CSI codebook size $N=\{64, 128, 256\}$.  }\label{Fig:CSIRateVsDoppler}
\end{figure}

Fig.~\ref{Fig:CSIRateVsDoppler} plots the normalized CSI source bit rate $R_sT$ in \eqref{Eq:AvRate:Erg} versus normalized Doppler shift $f_DT$ on a logarithmic scale. The size of the CSI codebook is $N = \{64, 128, 256\}$. As observed from Fig.~\ref{Fig:CSIRateVsDoppler}, $R_sT$ increases linearly with $f_DT$. Moreover, the increasing rate differs for different values of $N$ as reflected on the relative shifts between curves in Fig.~\ref{Fig:CSIRateVsDoppler}. This rate is higher for  larger $N$ and vise versa. These agree with the intuition that faster fading and finer CSI quantization generate CSI at a higher rate.

\begin{figure}
\centering
  \includegraphics[width=9.5cm]{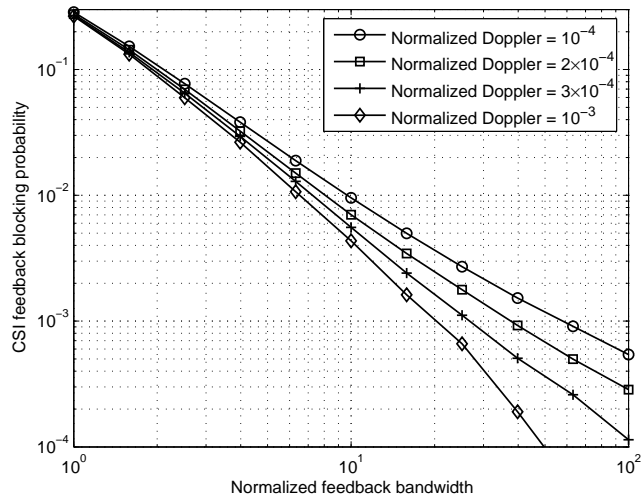}\\
  \caption{Feedback outage probability versus the CSI feedback bit rate normalized by the CSI source bit rate; the normalized Doppler shift is $f_DT = \{10^{-4},  2\times 10^{-4}, 3\times 10^{-4}, 10^{-3}\}$, and the CSI codebook size is $N=128$.} \label{Fig:OvProbVsBW}
\end{figure}

Fig.~\ref{Fig:OvProbVsBW} plots feedback outage probability $P_o$ versus the normalized CSI feedback bit rate $R_f/R_s$.  The normalized Doppler shift is $f_DT = \{10^{-4},  2\times 10^{-4}, 3\times 10^{-4}, 10^{-3}\}$, and the CSI codebook size is $N=128$. As observed from Fig.~\ref{Fig:OvProbVsBW}, for $R_f/R_s$ of practical interest (e.g. smaller than $6$), $P_o$ is insensitive to a change on $f_DT$. As observed from Fig.~\ref{Fig:OvProbVsBW}, $R_f/R_s$ must be larger than $2$ to keep $P_o$ reasonably low (e.g. smaller than $0.1$). By using Fig.~\ref{Fig:OvProbVsBW} and Fig.~\ref{Fig:CSIRateVsDoppler}, the CSI feedback bit rate can be obtained for a target feedback outage probability and a Doppler shift.

\subsection{Feedback Delay and Feedback Throughput Gain}\label{Section:Numerical:FbDelay}

\begin{figure}
\centering
  \includegraphics[width=9.5cm]{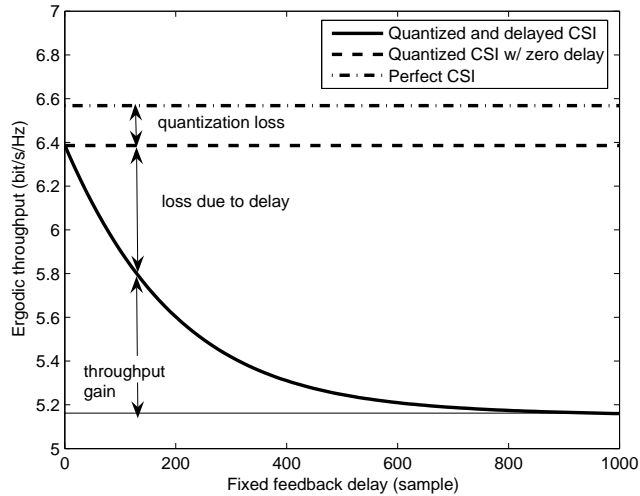}\\
  \caption{Comparison of  the effects of CSI quantization and feedback delay on the ergodic throughput of a transmit-beamforming system. Three cases of CSI feedback are considered, namely 1) perfect, 2) quantized, and 3) quantized and delayed CSI feedback. The number of antennas in an array is $L=4$, the SNR is $10$ dB, the normalized Doppler shift is $f_DT=10^{-3}$, and the quantizer codebook size is $N=128$.}\label{Fig:CmpMetric}
\end{figure}

Fig.~\ref{Fig:CmpMetric} compares the effects of CSI quantization and feedback delay on the ergodic throughput of a beamforming system. To be specific, the ergodic throughput in \eqref{Eq:ErgCap:a} is plotted against fixed feedback delay, which accounts for both CSI quantization and feedback delay. For simplicity, the variable delay due to the feedback protocol is omitted but addressed in the sequel. In addition, several  capacity metrics discussed in Section~\ref{Section:FBCompress:Cap} are illustrated in Fig.~\ref{Fig:CmpMetric}. For this figure, the number of antennas in an array is $L=4$, the SNR is $10$ dB, the normalized Doppler shift is $f_DT=10^{-3}$, and the quantizer codebook size is $N=128$. The quantization loss of $0.2$ b/s/Hz is small as also observed in \cite{LovHeaETAL:GrasBeamMultMult:Oct:03,MukSabETAL:BeamFiniRateFeed:Oct:03}. This loss can be controlled by adjusting the codebook size $N$. The quantization loss is smaller than the capacity loss due to feedback delay as it exceeds about $30$ samples. In other words, large feedback delay is a dominant source of CSI inaccuracy. Finally, the feedback throughput gain decreases exponentially with feedback delay. Thus, it is important to constrain the delay for retaining the capacity gain of transmit-beamforming.

\begin{figure}
    \centering
\hspace{-40pt}\subfigure[Normalized Doppler shift $f_DT=10^{-4}$]{  \includegraphics[width=9.5cm]{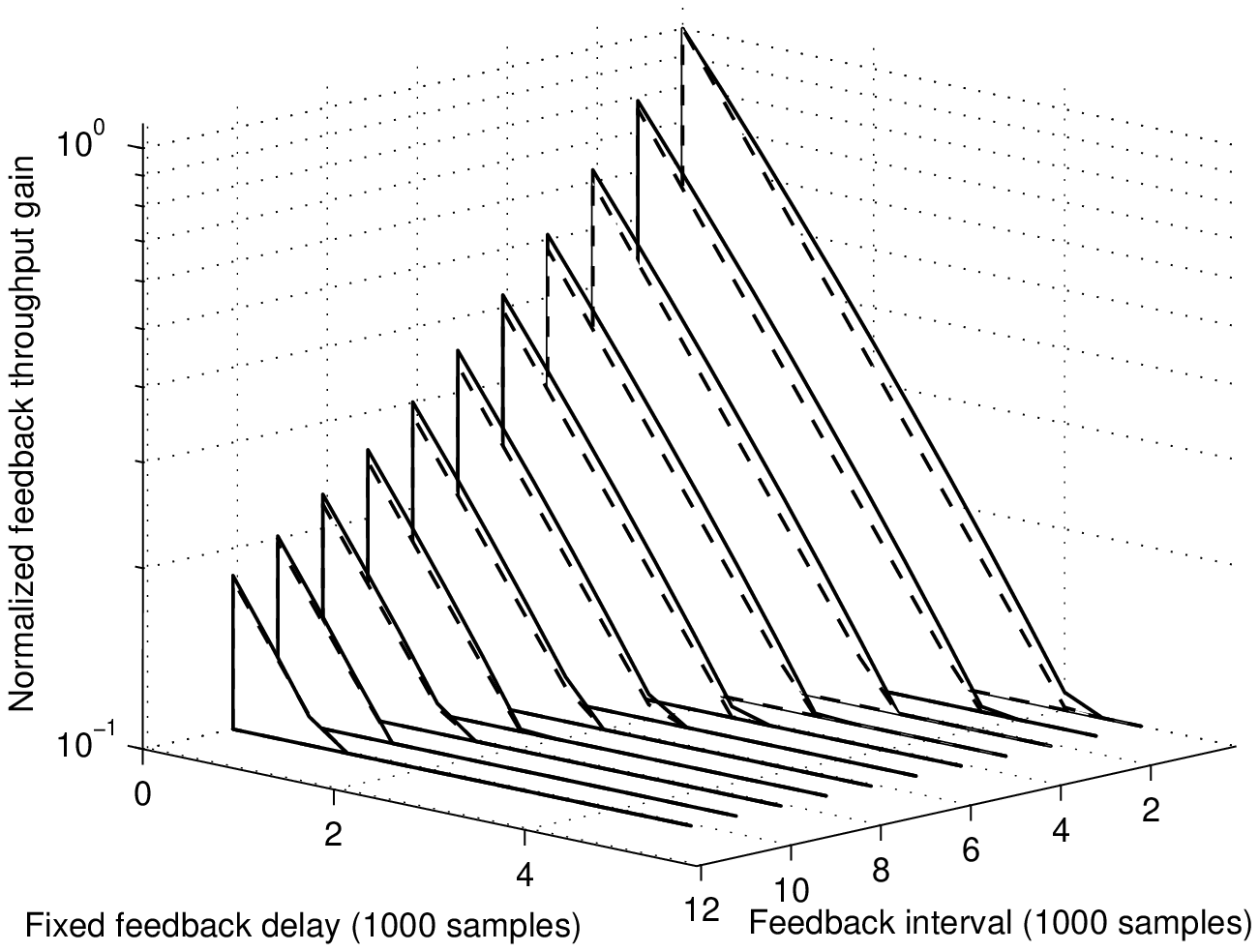}}\hspace{-30pt}
\subfigure[Normalized Doppler shift $f_DT=10^{-3}$]{  \includegraphics[width=9.5cm]{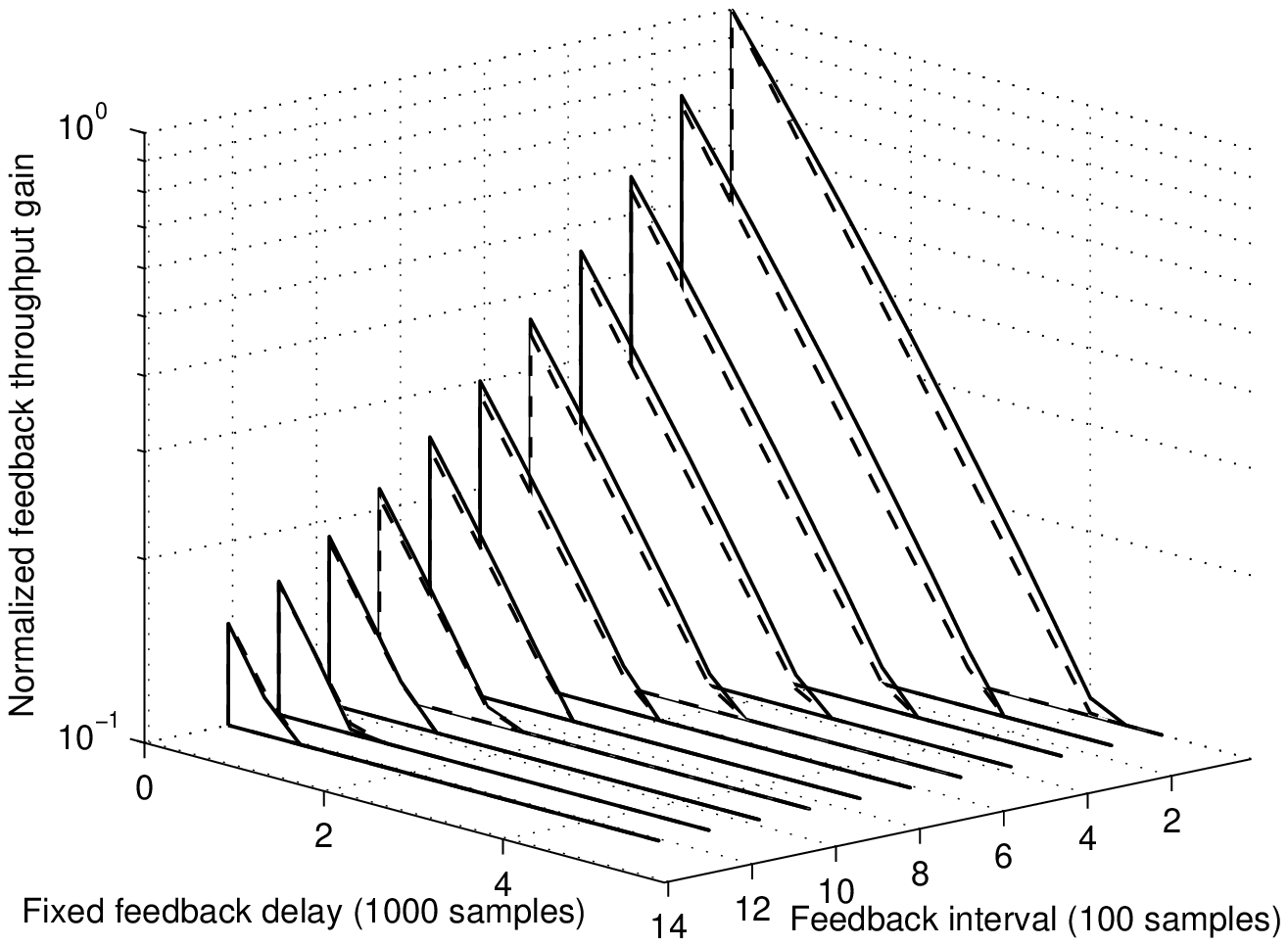}\hspace{-40pt}}\\
  \caption{The normalized feedback throughput gain and its approximation versus different combinations of the fixed and protocol delays; the normalized Doppler shift is (a) $f_DT=10^{-4}$ and (b) $f_DT=10^{-3}$, and the CSI codebook size is $N=128$.}\label{Fig:FbCapGainVsDly}
\end{figure}

Fig.~\ref{Fig:FbCapGainVsDly} compares the normalized feedback throughput gain $\Delta \bar{R}(D, K)$ and its approximation in \eqref{Eq:FbCapGain:Approx} for different combinations of the fixed feedback delay $D$ and the feedback interval $K$ both in samples.  Fig.~\ref{Fig:FbCapGainVsDly}(a) and Fig.~\ref{Fig:FbCapGainVsDly}(b) correspond to the normalized Doppler shift of $f_DT = 10^{-4}$ and $10^{-3}$, respectively. The CSI codebook size is $N=128$ and the SNR is 10 dB. The exact and the approximated values of $\Delta \bar{R}(D, K)$ are plotted by using solid and dashed lines, respectively. As observed from Fig.~\ref{Fig:FbCapGainVsDly}, $\Delta \bar{R}(D, K)$ is accurately approximated by \eqref{Eq:FbCapGain:Approx}. This allows the use of \eqref{Eq:FbCapGain:Approx}  for computing allowable feedback delay. Moreover, $\Delta \bar{R}(D, K)$ is observed to decrease with $D$ closely following the exponential rate as specified in Proposition~\ref{Prop:FbCapGain}. This rate grows with increasing normalized Doppler shift. This observation agrees with the intuition that the effect of CSI feedback delay  on throughput is more significant for fast fading and vise versa.

\subsection{Feedback Compression}\label{Section:Numerical:FbCompress}
Fig.~\ref{Fig:FbComp} plots the feedback compression ratio $\Delta R_f$ in \eqref{Eq:CompRatio} for different combinations of normalized Doppler shift $f_DT$ and the codebook size $N$. Fig.~\ref{Fig:FbComp}(a) and (b) correspond to the normalized CSI feedback bit rate of $R_f/R_s = 3$ and $5$, respectively. The threshold for truncating low-probability channel state transitions is $\epsilon = 0.1$ (cf. Section~\ref{Section:FBCompress:Algo}). As observed from Fig.~\ref{Fig:FbComp}, the feedback compression ratio $\Delta R_f$ is independent of normalized Doppler shift $f_DT$ for the following reason. Roughly speaking, a given value of $R_f/R_s$ fixes the degree of temporal correlation in feedback CSI, and thereby results in a constant $\Delta R_f$ independent of $f_DT$. By comparing Fig.~\ref{Fig:FbComp}(a) and Fig.~\ref{Fig:FbComp}(b), the feedback compression ratio $\Delta R_f$ increases as the normalized CSI feedback bit rate $R_f/R_s$ rises.  For example, for the codebook size of $N=128$,  $\Delta R_f$ is about 15\% for $R_f/R_s=3$, and rises to about 22\% for $R_f/R_s=5$. The reason for the above observation is that a larger value for $R_f/R_s$ strengthens temporal correlation in feedback CSI and thereby increases $\Delta R_f$. Therefore, there exists a trade-off between the feedback compression efficiency and the CSI feedback bit rate.

\setlength{\abovecaptionskip}{-0pt}

\subsection{Design Example}\label{Section:Numerical:DesignEx}
\begin{figure}
\centering
  \includegraphics[width=15cm]{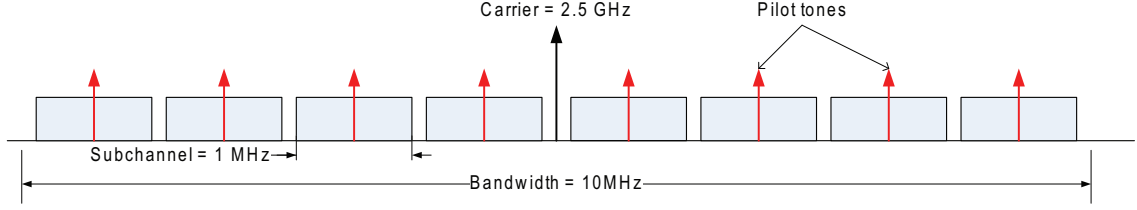}\\
  \caption{Spectrum of an OFDMA system at the $2.5$GHz band. The $10$MHz bandwidth is divided into $8$ subchannels of $1$ MHz. Each subchannel is assigned to one user.   }\label{Fig:OFDMA}
\end{figure}

This section presents a design example that demonstrates the joint application of the results in this paper for designing limited feedback beamforming. This example targets a broadband MIMO downlink system employing transmit-beamforming and \emph{orthogonal frequency division multiple access} (OFDMA) \cite{AndrewsBook:WiMax:07}, which have been included in both the IEEE 802.16 (WiMax) \cite{IEEE802-16e} and the 3GPP-LTE \cite{3GPP-LTE} standards. As illustrated in Fig.~\ref{Fig:OFDMA}, downlink OFDMA partitions a piece of radio spectrum into frequency slots, called \emph{subchannels}, and assigns them to different users for receiving data from a base station. For simplicity, scheduling is omitted in this example and users assigned to subchannels are assumed to be fixed regardless of their CSI. Modulated using \emph{orthogonal frequency division multiplexing} (OFDM), subchannels are fully decoupled  and each contains a number of orthogonal finer frequency channels, called \emph{subcarriers}. Based on limited feedback, transmit beamforming is performed on each subcarrier to increase its data rate. Assume that the narrow-band MIMO channels of different subcarriers in the same subchannel are approximately identical. Thus, only one CSI feedback link is required for each sub-channel. CSI is estimated at subscriber units by using pilot tones located at the centers of subchannels.

\begin{figure}
\centering
\hspace{-50pt}\subfigure[Normalized CSI feedback bit rate $R_f/R_s= 3$]{   \includegraphics[width=9.5cm]{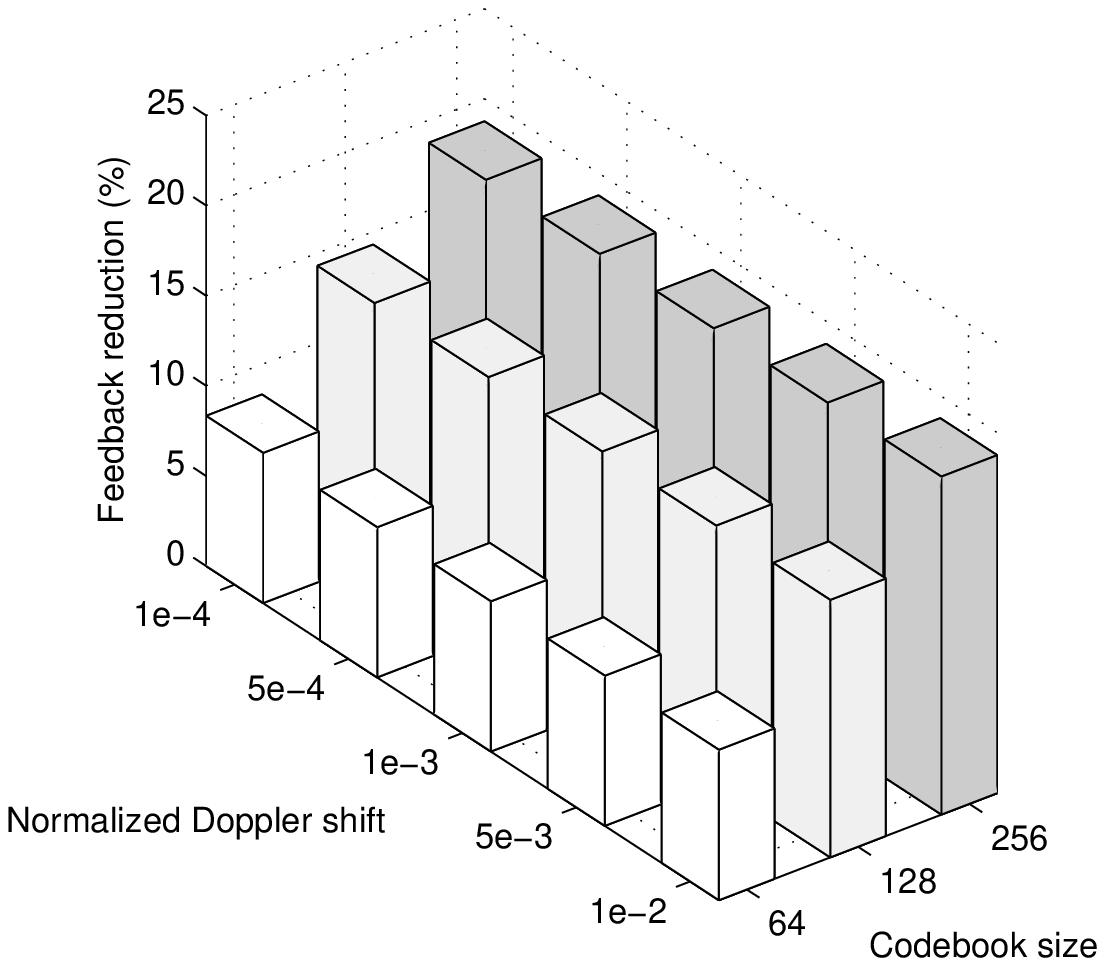}\hspace{-50pt}}
\subfigure[Normalized CSI feedback bit rate $R_f/R_s= 5$]{   \includegraphics[width=9.5cm]{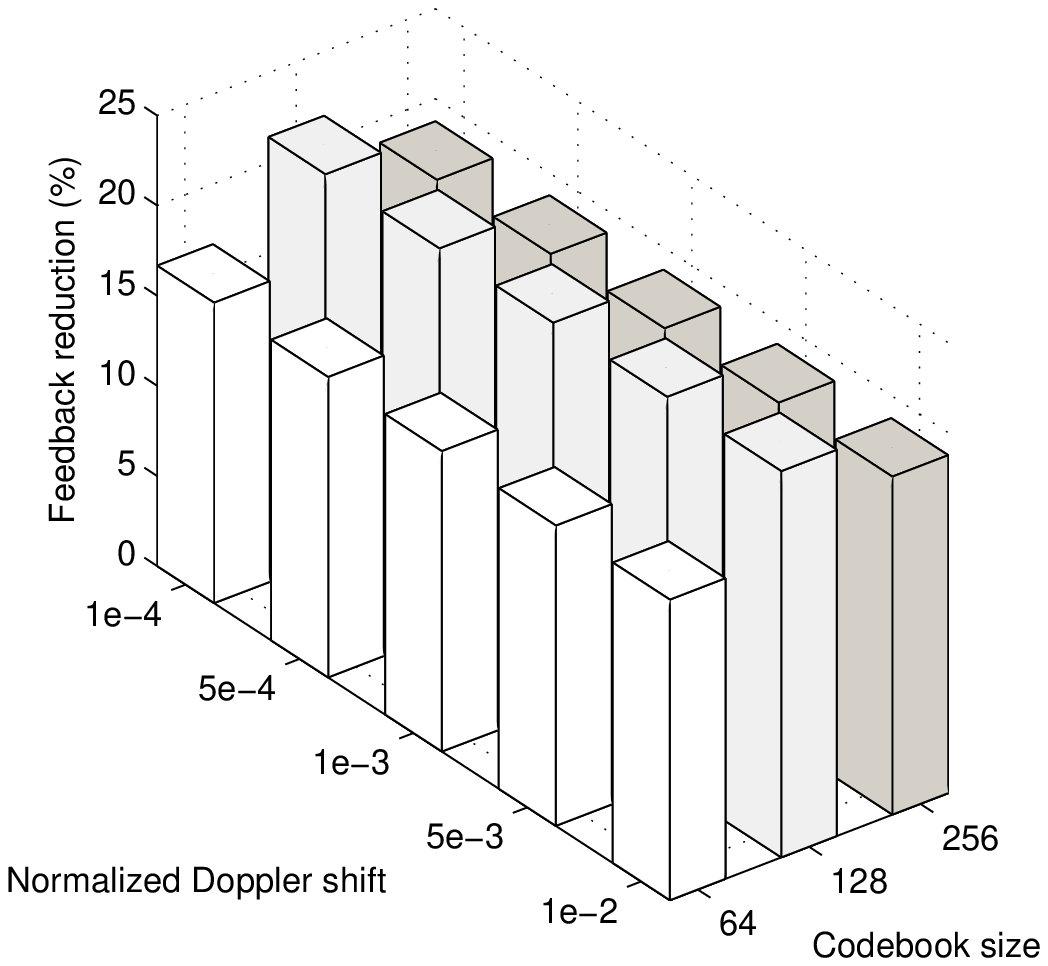}\hspace{-50pt}}\\
  \caption{Feedback compression ratio versus different combinations of Doppler shift and the CSI codebook size; the normalized feedback bit rate is (a) $R_f/R_s = 3$ and (b) $R_f/R_s = 5$, the threshold for truncating low-probability channel state transitions is $\epsilon = 0.1$. }\label{Fig:FbComp}
\end{figure}

The design specifications are summarized as follows. The spectrum is $10$ MHz at the $2.5$ GHz frequency band to be used for WiMax implementation. As shown in Fig.~\ref{Fig:OFDMA}, this spectrum consists of $8$ subchannels, each has $1$ MHz for data transmission. The remaining $2$ MHz bandwidth accounts for pilot and guard subcarriers. At each scheduled subscriber unit, the estimated CSI is sampled at $1$ MHz equal to the data symbol rate. Therefore, the sampling interval is $T = 1$ microsecond. The size of the codebook used for quantizing feedback CSI is $N=128$. The normalized feedback throughput gain in \eqref{Eq:FbCapGain:Approx} is required to be $\{50, 60, 70\}\%$. The mobility is up to $45$ km/h. The feedback outage probability is constrained as $ P_o \leq 6\times 10^{-2}$. These design requirements are also summarized in Table~\ref{Tab:Design}.

\begin{table}
  \centering
    \caption{Design example of a limited feedback beamforming system}\label{Tab:Design}
  \begin{tabular}{rlrl}
    \hline
    \multicolumn{4}{c}{{\bf Specifications}} \\
    Bandwidth:              & $10$ MHz          & Subchannel:               & $1$ MHz       \\
    Antenna array:          & $N_t=N_r = 4$     & Carrier frequency:         & $f_c = 2.5$ GHz \\
    Symbol rate: & $f_s = 1$ MHz                & CSI    codebook size:       & $N=128$ \\
    Normalized feedback throughput gain: & $\{50, 60, 70\}\%$     & Feedback overflow probability: & $P_o \leq 6\times 10^{-2}$\\
    Mobility:               &$V \leq 45$ km/h   & & \\
    \hline
    \multicolumn{4}{c}{
    \begin{tabular}{rcccccc}
    \multicolumn{7}{c}{\bf Computed Parameters}\\
    \multicolumn{1}{r}{{}}&\multicolumn{3}{l}{(w/o compression)}& \multicolumn{3}{l}{(w/ compression)}\\
    Feedback throughput gain (b/s/Hz)   & 0.60          & 0.72      & 0.84 & 0.54 & 0.65 & 0.76\\
    (Normalized)                      &  50\%         & 60\%      & 70\% & 45\% & 54\% & 63\% \\
    Throughput gain/subchannel (Mb/s)        & $0.6$         & $0.72$    & $0.84$ & $0.54$ & $0.65$ & $0.76$\\
    Total throughput gain (Mb/s)            & $4.8$         & $5.76$    & $6.72$  & $4.32$ & $5.20$ & $6.08$\\
    CSI feedback rate/subchannel (kbits/s)    & 21           & 21       & 21 & 17.9 & 17.9 & 17.9\\
    Sum feedback rate (kbits/s)    & 168           & 168       & 168 & 143.2 & 143.2 & 143.2\\
    (Normalized)                      & 3           & 3           & 3     & 2.6 & 2.6 & 2.6\\
    Maximum fixed delay (ms)          & 0.99 & 0.69 & 0.43 &  0.99 & 0.69 & 0.43\\

  \end{tabular}
}\\    \hline
\end{tabular}
\end{table}

Based on above design specifications, the system parameters are computed as follows. Consider  CSI feedback for an arbitrary subchannel. The Doppler shift for the maximum speed of 45 km/h is obtained as $f_D = 104$ Hz \cite{GoldsmithBook:WirelessComm:05},  and normalized by the CSI sampling rate to be $f_DT = 10^{-4}$. Based on the this result and the feedback outage constraint, the corresponding normalized feedback bit rate is obtained from Fig.~\ref{Fig:OvProbVsBW} to be $R_f/R_s = 3$.  By using Fig.~\ref{Fig:CSIRateVsDoppler} and for the symbol rate of 1 MHz and the codebook size of $N=128$, the CSI source bit rate is obtained as $R_s = 3$ kbit/s and hence the CSI feedback bit rate is $R_f = 21$ Kbit/s. It follows that the feedback interval is $KT = 0.33$ millisecond. From Fig.~\ref{Fig:FbComp}, the feedback bit rate can be compressed by about $14\%$ for $N=128$ and $f_DT = 10^{-4}$. Therefore, the compressed CSI feedback bit rate is $\hat{R}_f = 17.9$ kbit/s. By combining all subchannels, the sum feedback rate is $168$ Kbit/s without compression and $143.2$ Kbit/s with compression.


Next, the maximum fixed delay for each CSI feedback link is computed as follows.  Corresponding to the codebook size of $N=128$ and the normalized Doppler shift of $f_DT = 10^{-4}$, the eigenvalue defined in Proposition~\ref{Prop:FbCapGain} is computed to be $\sqrt{\lambda} = 0.9994$ by using the stochastic matrix of the channel state Markov chain. By substituting this result into \eqref{Eq:FbCapGain:Approx}, the maximum fixed feedback delay is obtained as $D_{\max} = \{992, 688, 431\}$ samples, corresponding to the normalized feedback throughput gain of $\{50, 60, 70\}\%$, respectively. Equivalently, $D_{\max} = \{0.99, 0.69, 0.43\}$ millisecond since the symbol rate is $1$ MHz. Last, the maximum feedback throughput gain for delay-free CSI feedback is computed to be $1.2$ bit/s/Hz by using \eqref{Eq:ErgCap:NoDly}. Therefore, the required normalized feedback throughput gains of $\{50, 60, 70\}\%$ are converted into the actual values of $\{0.6, 0.72, 0.84\}$ bit/s/Hz. This corresponds to the increase in throughput equal to $\{0.6, 0.72, 0.84\}$ Mbit/s per subchannel, and $\{4.8, 5.76, 6.72\}$ Mbit/s over the whole $10$ MHz spectrum. The system parameters computed above are summarized in Table~\ref{Tab:Design}.

A few remarks are given on the designed parameters as given in Table~\ref{Tab:Design}. First, the required CSI feedback bit rates in Table~\ref{Tab:Design} are small due to relatively low mobility and large coherence bandwidth ($\geq 1$ MHz).  Second, increasing the feedback throughput gain decreases the maximum allowable fixed delay. In other words, higher throughput requires faster signal processing and shorter transmission distance. Finally, the CSI feedback reduction by compression is moderate since the normalized CSI feedback bit rate and hence the correlation in feedback CSI are small.

\section{Conclusion}
\label{Section:Conclusion} We have designed limited feedback beamforming systems over  temporally-correlated channels. The quantized CSI has been modeled as a first-order Markov chain to provide an analytical tool. This model has been validated by simulation. Based on this model, the CSI source bit rate has been derived as a function of the Markov chain probabilities. Moreover, adopting a periodic feedback protocol, the CSI feedback bit rate supported by the feedback channel has been obtained under a feedback outage constraint. Next, an upper bound on the feedback throughput gain has been derived as a function of the CSI feedback delay and feedback interval. Last, a feedback compression algorithm has been proposed for reducing the CSI feedback bit rate by exploiting temporal correlation in feedback CSI. Simulation results have been generated based on spatially i.i.d. Rayleigh fading and the Clarke's fading model. From these results, the CSI source bit rate has been observed to increase linearly with Doppler shift. The ratio between CSI feedback and source bit rates has been found to be insensitive to a change on Doppler shift. The theoretical upper-bound on the feedback throughput gain has been shown to be tight, confirming the exponential decay rate of the feedback throughput gain with feedback delay. Moreover, the proposed feedback compression algorithm has been observed to achieve a compression ratio more than $20\%$. Finally, an design example for a limited feedback beamforming system has been presented, which demonstrates the usefulness of this work for practical applications.

This paper opens several issues for future investigation. First, the results of this work focusing on limited feedback beamforming can be readily extended to other types of limited feedback systems such as precoding, multiuser MIMO, and adaptive modulation and power control. Specifically, the same approach based on Markov chain theory as used in this paper can be applied by properly defining Markov chain states based the type of limited feedback system. Second, channel estimation errors omitted in this paper should be considered in future work. In particular, the training data for CSI estimation should be jointly designed with feedback CSI for optimizing system performance and minimizing total overhead. Third, a rigorous study of channel prediction for coping with feedback delay is useful for designing practical limited feedback systems. Fourth, it is important to investigate the required order of the channel state Markov chain for accurately modeling CSI for different kinds of channel distributions such as Rayleigh and Rician and temporal correlation. Finally, the time variation of Doppler shift typical in the practice should be accounted for in future work. In particular, the time-varying Doppler shift changes system parameters including CSI feedback bit and allowable feedback delay.

\section*{Acknowledgement}
The authors thank Dr. Bishwarup Mondal for helpful discussion in the early stage of this work.

\appendix

\subsection{Proof of Proposition~\ref{Prop:FbBW}}\label{App:FbBW}
Consider the variation of the channel state in the $\tilde{n}$th feedback interval $(\tilde{n}-1)K+1\leq n \leq  \tilde{n}K$ and let $Z_{\tilde{n}}$ denote the number of channel-state transitions in this interval. This does not compromise the generality because the channel state Markov chain is stationary. Let $\ell_0\in\mathcal{I}$ denotes the initial channel state for the $\tilde{n}$th feedback interval, thus $I_{(\tilde{n}-1)K}=\ell_0$. Conditioned on this initial state, the probability for each sample path of the channel state in the $\tilde{n}$th feedback interval, denoted as $\ell_1\rightarrow \ell_2\cdots\rightarrow \ell_K$, can be expressed in terms of channel state transition probabilities as $\Pr(\ell_1\rightarrow \ell_2\cdots\rightarrow \ell_K\mid \ell_0)= P_{\ell_{K}\ell_{K-1}}P_{\ell_{K-1}\ell_{K-2}} \cdots P_{\ell_1\ell_0}$. It follows that the conditional probability of no transition is
\begin{equation}\label{Eq:CondPr:a}
\Pr(Z_{\tilde{n}} = 0\mid I_{(\tilde{n}-1)K}=\ell_0) = \left[\bP^K\right]_{\ell_0\ell_0}.
\end{equation}
The conditional probability of single channel-state transition is obtained by grouping all sample paths containing only one transition. Thus,
\begin{equation}\label{Eq:CondPr:b}
\Pr(Z_{\tilde{n}} = 1 \mid I_{(\tilde{n}-1)K}=\ell_0) = \sum_{k=1}^K\sum_{\substack{m=1\\ m\neq \ell_0}}^N[\bP^{K-k}]_{mm}P_{m\ell_0}[\bP^{k-1}]_{\ell_0\ell_0} .
\end{equation}
From \eqref{Eq:CondPr:a} and \eqref{Eq:CondPr:b}
\begin{eqnarray}
\Pr(Z > 1 )  &=& 1- \sum_{\ell=1}^N\left[\Pr(Z_{\tilde{n}} =0 \mid I_{(k-1)K}=\ell) + \Pr(Z_{\tilde{n}} =1\mid I_{(k-1)K}=\ell)\right]\pi_{\ell} \nn\\
&=& 1 - \sum_{\ell=1}^NP^K_{\ell\ell}\pi_{\ell} - \sum_{k=1}^K\sum_{\ell=1}^N\sum_{\substack{m=1\\ m\neq \ell}}^N P_{mm}^{K-k}P_{m\ell}P_{\ell\ell}^{k-1} \pi_{\ell}\nn
\end{eqnarray}
\begin{eqnarray}
&=& 1 - \sum_{\ell=1}^N P^K_{\ell\ell}\pi_{\ell} - \sum_{\ell=1}^N\sum_{\substack{m=1\\ m\neq \ell}}^N\frac{P_{mm}^K P_{m\ell}\pi_{\ell}}{P_{\ell\ell}}\sum_{k=1}^K\left(\frac{P_{\ell\ell}}{P_{mm}}\right)^k\nn\\
&=& 1 - \sum_{\ell=1}^N P^K_{\ell\ell}\pi_{\ell} - \sum_{\ell=1}^N\sum_{\substack{m=1\\ m\neq \ell}}^N\frac{(P_{mm}^K - P_{\ell\ell}^K)P_{m\ell}\pi_{\ell}}{P_{mm}-P_{\ell\ell}}.\nn
\end{eqnarray}
Thus, by combining the above equation and the feedback constraint, the desired result follows.

\subsection{Proof of Corollary~\ref{Cor:FbBw}}\label{App:FbBw:Cor}
The left-hand-size of \eqref{Eq:FbConstraint:a} attains its maximum value of one at $K=1$. Thus, the result for $\delta\rightarrow 0$ follows.

Next, consider the case of $\delta \rightarrow 1$ . An upper bound on $K$ is obtained by considering the following upper-bound on the left-hand-size of \eqref{Eq:FbConstraint:a}
\begin{eqnarray}
\sum_{\ell=1}^NP^K_{\ell\ell}\pi_{\ell} + \sum_{m\neq \ell}\frac{(P_{mm}^K - P_{\ell\ell}^K)P_{m\ell}\pi_{\ell}}{P_{mm}-P_{\ell\ell}} &=& \sum_{\ell=1}^NP^K_{\ell\ell}\pi_{\ell} + \sum_{m\neq \ell}\frac{[\max(P_{mm}, P_{\ell\ell})^K - \min(P_{mm}, P_{\ell\ell})^K]P_{m\ell}\pi_{\ell}}{|P_{mm}-P_{\ell\ell}|}\nn\\
&\leq& \sum_{\ell=1}^NP^K_{\ell\ell}\pi_{\ell} + \sum_{m\neq \ell}\frac{(\max_a P_{aa})^K P_{m\ell}\pi_{\ell}}{|P_{mm}-P_{\ell\ell}|}\nn\\
&\leq& (\max_a P_{aa})^K\left(1 + \sum_{m\neq \ell}\frac{P_{m\ell}\pi_{\ell}}{|P_{mm}-P_{\ell\ell}|}\right). \nn
\end{eqnarray}
Therefore, the following constraint is the relaxation of that in \eqref{Eq:FbConstraint:a}
\begin{equation}\label{Eq:FbConstraint:b}
(\max_a P_{aa})^K\left(1 + \sum_{m\neq \ell}\frac{P_{m\ell}\pi_{\ell}}{|P_{mm}-P_{\ell\ell}|}\right)\geq 1-\delta.
\end{equation}
From the above constraint, an upper bound on $K$, denoted as $K^+$, is obtained as
\begin{equation}\label{Eq:K:UB}
K^+ = \frac{\log(1-\delta) - \log\left(1 + \sum_{m\neq \ell}\frac{P_{m\ell}\pi_{\ell}}{|P_{mm}-P_{\ell\ell}|}\right)}{\log\left(\max_a P_{aa}\right)}.
\end{equation}
To obtain an lower bound for $K$, the left-hand-size of \eqref{Eq:FbConstraint:a} is bounded from below as
\begin{eqnarray}
\sum_{\ell=1}^NP^K_{\ell\ell}\pi_{\ell} + \sum_{m\neq \ell}\frac{(P_{mm}^K - P_{\ell\ell}^K)P_{m\ell}\pi_{\ell}}{P_{mm}-P_{\ell\ell}} &\geq& \sum_{\ell=1}^NP^K_{\ell\ell}\pi_{\ell} \geq \left(\min_{a}P_{aa}\right)^K.
 \end{eqnarray}
Thus, the constraint $\left(\min_{a}P_{aa}\right)^K \geq 1-\delta$ is more stringent than that in \eqref{Eq:FbConstraint:a}. Based on this constraint, a lower-bound on $K$, denoted as $K^-$, is obtained as
\begin{equation}\label{Eq:K:LB}
K^- = \frac{\log(1-\delta)}{\log(\min_a P_{aa})}.
\end{equation}
Combining \eqref{Eq:K:LB} and \eqref{Eq:K:UB} gives the desired result.

\subsection{Proof of Lemma~\ref{Lem:ErgCap}}\label{App:ErgCap}

From \eqref{Eq:ErgCap}
\begin{eqnarray}
 R &=& \sum_{\ell=1}^N\sum_{m=1}^N\E\left[\log_2(1+P\|\bH_n \mathcal{Q}_f(\bH_{n-d})\|^2)\mid \bH_n\in\mathcal{V}_{\ell}, \bH_{n-d}\in\mathcal{V}_m\right]\Pr(\bH_n\in\mathcal{V}_{\ell}, \bH_{n-d}\in\mathcal{V}_m)\nn\\
&=& \sum_{\ell=1}^N\sum_{m=1}^N\E\left[\log_2(1+P\|\bH_n \bv_m\|^2)\mid \bH_n\in\mathcal{V}_{\ell}, \bH_{n-d}\in\mathcal{V}_m\right]\Pr(\bH_n\in\mathcal{V}_{\ell}, \bH_{n-d}\in\mathcal{V}_m)\nn\\
&=& \sum_{\ell=1}^N\sum_{m=1}^NR_{\ell m}\Pr(I_n=\ell, I_{n-d}=m)\nn
\end{eqnarray}
where $R_{\ell m}$ is defined in \eqref{Eq:ErgCap:Comp},  $\bv_m$ is the $m$th member of the codebook $\mathcal{F}$, and $\mathcal{V}_{\ell}$ is the Voronoi cell.

\subsection{Proof of Corollary~\ref{Cor:ErgCap:SpecCases}}\label{App:ErgCap:SpecCases}
By substituting $D=0$ into \eqref{Eq:ErgCap:a}
\begin{equation}
R(0) = \sum_{\ell=1}^N\E\left[\log_2(1+P\|\bH_n \mathcal{Q}_f(\bH_n)\|^2)\mid \bH_n\in\mathcal{V}_{\ell}\right]\pi_{\ell}\nn.
\end{equation}
Given that $\cup_{\ell}\mathcal{V}_{\ell} =\mathds{C}^{N_r\times N_t}$, the result in \eqref{Eq:ErgCap:NoDly} follows from the above equation. Next, by substituting \eqref{Eq:FSMC:Property} into \eqref{Eq:ErgCap:a}, the first equality in \eqref{Eq:ErgCap:InfDly} is obtained. The second inequality in \eqref{Eq:ErgCap:InfDly} follows from \eqref{Eq:MarkovPara}, \eqref{Eq:ErgCap:Comp} and the fact that $\cup_{\ell}\mathcal{V}_{\ell} =\mathds{C}^{N_r\times N_t}$.

\subsection{Proof of Proposition~\ref{Prop:FbCapGain}}\label{Appendix:GainRate}
From \eqref{Eq:FbCapGain:Def}, the feedback throughput gain is upper-bounded as
\begin{eqnarray}
R(D)  &\leq& \sum_{\ell=1}^{N}\sum_{m=1}^{N}R_{\ell m}\left|\left[\bP^D\right]_{\ell m}-\pi_{\ell}\right|\pi_m\nn\\
      &\leq& \sum_{m=1}^{N}\pi_m\left(\max_{\ell}R_{\ell m}\right)\sum_{\ell = 1}^N\left|\left[\bP^D\right]_{\ell m}-\pi_{\ell}\right|\label{Eq:App:a}
\end{eqnarray}
The desired result is obtained from \eqref{Eq:App:a} and Lemma~\ref{Lem:RedRate}.

\bibliographystyle{ieeetr}


\end{document}

%% file: header.tex
\newtheorem{theorem}{\it Theorem}
\newtheorem{acknowledgement}[theorem]{Acknowledgement}
\newtheorem{axiom}[theorem]{Axiom}
\newtheorem{case}[theorem]{Case}
\newtheorem{claim}[theorem]{Claim}
\newtheorem{conclusion}[theorem]{Conclusion}
\newtheorem{condition}[theorem]{Condition}
\newtheorem{conjecture}[theorem]{Conjecture}
\newtheorem{criterion}[theorem]{Criterion}
\newtheorem{definition}[theorem]{Definition}
\newtheorem{example}[theorem]{Example}
\newtheorem{exercise}[theorem]{Exercise}
\newtheorem{lemma}{Lemma}
\newtheorem{corollary}{Corollary}
\newtheorem{notation}[theorem]{Notation}
\newtheorem{problem}[theorem]{Problem}
\newtheorem{proposition}{Proposition}
\newtheorem{solution}[theorem]{Solution}
\newtheorem{summary}[theorem]{Summary}
\newtheorem{assumption}{Assumption}
\newtheorem{examp}{\bf Example}
\newtheorem{probform}{\bf Problem}
\def\remark{{\noindent \bf Remark:\hspace{0.5em}}}

\def\qed{$\Box$}
\def\QED{\mbox{\phantom{m}}\nolinebreak\hfill$\,\Box$}
\def\proof{\noindent{\emph{Proof:} }}
\def\poof{\noindent{\emph{Sketch of Proof:} }}
\def
\endproof{\hspace*{\fill}~\qed
\par
\endtrivlist\unskip}
\def\endproof{\hspace*{\fill}~\qed\par\endtrivlist\vskip3pt}

\def\E{\mathbb{E}}
\def\eps{\varepsilon}
\def\phi{\varphi}
\def\Lsp{{\boldsymbol L}}
\def\Bsp{{\boldsymbol B}}
\def\lsp{{\boldsymbol\ell}}
\def\Ltsp{{\Lsp^2}}
\def\Lpsp{{\Lsp^p}}
\def\Linsp{{\Lsp^{\infty}}}
\def\LtR{{\Lsp^2(\Rst)}}
\def\ltZ{{\lsp^2(\Zst)}}
\def\ltsp{{\lsp^2}}
\def\ltZt{{\lsp^2(\Zst^{2})}}
\def\ninN{{n{\in}\Nst}}
\def\oh{{\frac{1}{2}}}
\def\grass{{\cal G}}
\def\ord{{\cal O}}
\def\dist{{d_G}}
\def\conj#1{{\overline#1}}
\def\ntoinf{{n \rightarrow \infty }}
\def\toinf{{\rightarrow \infty }}
\def\tozero{{\rightarrow 0 }}
\def\trace{{\operatorname{trace}}}
\def\ord{{\cal O}}
\def\UU{{\cal U}}
\def\rank{{\operatorname{rank}}}
\def\acos{{\operatorname{acos}}}

\def\SINR{\mathrm{SINR}}
\def\SNR{\mathrm{SNR}}
\def\SIR{\mathrm{SIR}}

\setcounter{page}{1}

\newcommand{\eref}[1]{(\ref{#1})}
\newcommand{\fig}[1]{Fig.\ \ref{#1}}

\def\bydef{:=}
\def\ba{{\mathbf{a}}}
\def\bb{{\mathbf{b}}}
\def\bc{{\mathbf{c}}}
\def\bd{{\mathbf{d}}}
\def\bee{{\mathbf{e}}}
\def\bff{{\mathbf{f}}}
\def\bg{{\mathbf{g}}}
\def\bh{{\mathbf{h}}}
\def\bi{{\mathbf{i}}}
\def\bj{{\mathbf{j}}}
\def\bk{{\mathbf{k}}}
\def\bl{{\mathbf{l}}}
\def\bm{{\mathbf{m}}}
\def\bn{{\mathbf{n}}}
\def\bo{{\mathbf{o}}}
\def\bp{{\mathbf{p}}}
\def\bq{{\mathbf{q}}}
\def\br{{\mathbf{r}}}
\def\bs{{\mathbf{s}}}
\def\bt{{\mathbf{t}}}
\def\bu{{\mathbf{u}}}
\def\bv{{\mathbf{v}}}
\def\bw{{\mathbf{w}}}
\def\bx{{\mathbf{x}}}
\def\by{{\mathbf{y}}}
\def\bz{{\mathbf{z}}}
\def\b0{{\mathbf{0}}}

\def\bA{{\mathbf{A}}}
\def\bB{{\mathbf{B}}}
\def\bC{{\mathbf{C}}}
\def\bD{{\mathbf{D}}}
\def\bE{{\mathbf{E}}}
\def\bF{{\mathbf{F}}}
\def\bG{{\mathbf{G}}}
\def\bH{{\mathbf{H}}}
\def\bI{{\mathbf{I}}}
\def\bJ{{\mathbf{J}}}
\def\bK{{\mathbf{K}}}
\def\bL{{\mathbf{L}}}
\def\bM{{\mathbf{M}}}
\def\bN{{\mathbf{N}}}
\def\bO{{\mathbf{O}}}
\def\bP{{\mathbf{P}}}
\def\bQ{{\mathbf{Q}}}
\def\bR{{\mathbf{R}}}
\def\bS{{\mathbf{S}}}
\def\bT{{\mathbf{T}}}
\def\bU{{\mathbf{U}}}
\def\bV{{\mathbf{V}}}
\def\bW{{\mathbf{W}}}
\def\bX{{\mathbf{X}}}
\def\bY{{\mathbf{Y}}}
\def\bZ{{\mathbf{Z}}}

\def\mA{{\mathbb{A}}}
\def\mB{{\mathbb{B}}}
\def\mC{{\mathbb{C}}}
\def\mD{{\mathbb{D}}}
\def\mE{{\mathbb{E}}}
\def\mF{{\mathbb{F}}}
\def\mG{{\mathbb{G}}}
\def\mH{{\mathbb{H}}}
\def\mI{{\mathbb{I}}}
\def\mJ{{\mathbb{J}}}
\def\mK{{\mathbb{K}}}
\def\mL{{\mathbb{L}}}
\def\mM{{\mathbb{M}}}
\def\mN{{\mathbb{N}}}
\def\mO{{\mathbb{O}}}
\def\mP{{\mathbb{P}}}
\def\mQ{{\mathbb{Q}}}
\def\mR{{\mathbb{R}}}
\def\mS{{\mathbb{S}}}
\def\mT{{\mathbb{T}}}
\def\mU{{\mathbb{U}}}
\def\mV{{\mathbb{V}}}
\def\mW{{\mathbb{W}}}
\def\mX{{\mathbb{X}}}
\def\mY{{\mathbb{Y}}}
\def\mZ{{\mathbb{Z}}}

\def\cA{\mathcal{A}}
\def\cB{\mathcal{B}}
\def\cC{\mathcal{C}}
\def\cD{\mathcal{D}}
\def\cE{\mathcal{E}}
\def\cF{\mathcal{F}}
\def\cG{\mathcal{G}}
\def\cH{\mathcal{H}}
\def\cI{\mathcal{I}}
\def\cJ{\mathcal{J}}
\def\cK{\mathcal{K}}
\def\cL{\mathcal{L}}
\def\cM{\mathcal{M}}
\def\cN{\mathcal{N}}
\def\cO{\mathcal{O}}
\def\cP{\mathcal{P}}
\def\cQ{\mathcal{Q}}
\def\cR{\mathcal{R}}
\def\cS{\mathcal{S}}
\def\cT{\mathcal{T}}
\def\cU{\mathcal{U}}
\def\cV{\mathcal{V}}
\def\cW{\mathcal{W}}
\def\cX{\mathcal{X}}
\def\cY{\mathcal{Y}}
\def\cZ{\mathcal{Z}}
\def\cd{\mathcal{d}}
\def\Mt{M_{t}}
\def\Mr{M_{r}}
\def\O{\Omega_{M_{t}}}
\newcommand{\figref}[1]{{Fig.}~\ref{#1}}
\newcommand{\tabref}[1]{{Table}~\ref{#1}}

\newcommand{\var}{\mathrm{Var}}
\newcommand{\fb}{\tx{fb}}
\newcommand{\nf}{\tx{nf}}
\newcommand{\BC}{\tx{(bc)}}
\newcommand{\MAC}{\tx{(mac)}}
\newcommand{\Pout}{P_{\tx{out}}}
\newcommand{\nnn}{\nn\\}
\newcommand{\FB}{\tx{FB}}
\newcommand{\TX}{\tx{TX}}
\newcommand{\RX}{\tx{RX}}
\renewcommand{\mod}{\tx{mod}}
\newcommand{\m}[1]{\mathbf{#1}}
\newcommand{\td}[1]{\tilde{#1}}
\newcommand{\sbf}[1]{\scriptsize{\textbf{#1}}}
\newcommand{\stxt}[1]{\scriptsize{\textrm{#1}}}
\newcommand{\suml}[2]{\sum\limits_{#1}^{#2}}
\newcommand{\sumlk}{\sum\limits_{k=0}^{K-1}}
\newcommand{\eqhsp}{\hspace{10 pt}}
\newcommand{\tx}[1]{\texttt{#1}}
\newcommand{\Hz}{\ \tx{Hz}}
\newcommand{\sinc}{\tx{sinc}}
\newcommand{\tr}{\mathrm{tr}}
\newcommand{\diag}{\mathrm{diag}}
\newcommand{\MAI}{\tx{MAI}}
\newcommand{\ISI}{\tx{ISI}}
\newcommand{\IBI}{\tx{IBI}}
\newcommand{\CN}{\tx{CN}}
\newcommand{\CP}{\tx{CP}}
\newcommand{\ZP}{\tx{ZP}}
\newcommand{\ZF}{\tx{ZF}}
\newcommand{\SP}{\tx{SP}}
\newcommand{\MMSE}{\tx{MMSE}}
\newcommand{\MINF}{\tx{MINF}}
\newcommand{\RC}{\tx{MP}}
\newcommand{\MBER}{\tx{MBER}}
\newcommand{\MSNR}{\tx{MSNR}}
\newcommand{\MCAP}{\tx{MCAP}}
\newcommand{\vol}{\tx{vol}}
\newcommand{\ah}{\hat{g}}
\newcommand{\tg}{\tilde{g}}
\newcommand{\teta}{\tilde{\eta}}
\newcommand{\heta}{\hat{\eta}}
\newcommand{\uh}{\m{\hat{s}}}
\newcommand{\eh}{\m{\hat{\eta}}}
\newcommand{\hv}{\m{h}}
\newcommand{\hh}{\m{\hat{h}}}
\newcommand{\Po}{P_{\mathrm{out}}}
\newcommand{\Poh}{\hat{P}_{\mathrm{out}}}
\newcommand{\Ph}{\hat{\gamma}}
\newcommand{\mat}[1]{\begin{matrix}#1\end{matrix}}
\newcommand{\ud}{^{\dagger}}
\newcommand{\C}{\mathcal{C}}
\newcommand{\nn}{\nonumber}
\newcommand{\nInf}{U\rightarrow \infty}

%% file: MarkovLimFb_Arxiv.bbl
\begin{thebibliography}{10}

\bibitem{LoveHeathBook}
D.~J. Love and R.~W. Heath~Jr., ``Feedback techniques for {MIMO} channels,'' in
  {\em {MIMO} Antenna Technology for Wireless Communications}, (Boca Raton,
  FL), CRC Press Inc, 2006.

\bibitem{GoldsmithBook:WirelessComm:05}
A.~Goldsmith, {\em Wireless Communications}.
\newblock Cambridge University Press, 2005.

\bibitem{GerGra:VectQuanSignComp:92}
A.~Gersho and R.~M. Gray, {\em Vector Quantization and Signal Compression}.
\newblock Kluwer Academic Press, 1992.

\bibitem{LovHeaETAL:GrasBeamMultMult:Oct:03}
D.~J. Love, R.~W. Heath~Jr., and T.~Strohmer, ``Grassmannian beamforming for
  multiple-input multiple-output wireless systems,'' {\em IEEE Trans. on Info.
  Theory}, vol.~49, pp.~2735--47, Oct. 2003.

\bibitem{MukSabETAL:BeamFiniRateFeed:Oct:03}
K.~K. Mukkavilli, A.~Sabharwal, E.~Erkip, and B.~Aazhang, ``On beamforming with
  finite rate feedback in multiple antenna systems,'' {\em IEEE Trans. on Info.
  Theory}, vol.~49, pp.~2562--79, Oct. 2003.

\bibitem{Roh:EffFbMIMOParameter:2007}
J.~C. Roh and B.~D. Rao, ``Efficient feedback methods for {MIMO} channels based
  on parameterization,'' {\em IEEE Trans. on Wireless Communications}, vol.~6,
  pp.~282--292, Jan. 2007.

\bibitem{ChoiMondal:InterpPrecodSpaMuxMIMOOFDMLimFb:2006}
J.~Choi, B.~Mondal, and R.~W. Heath, ``Interpolation based unitary precoding
  for spatial multiplexing {MIMO}-{OFDM} with limited feedback,'' {\em IEEE
  Trans. on Sig. Proc.}, vol.~54, pp.~4730--40, Dec. 2006.

\bibitem{Xia:AchieveWelchBound:05}
P.~Xia, S.~Zhou, and G.~B. Giannakis, ``Achieving the {Welch} bound with
  difference sets,'' {\em IEEE Trans. on Info. Theory}, vol.~51, pp.~1900--07,
  May 2005.

\bibitem{LoveHeath:LimitedFeedbackPrecodOSTBC:05}
D.~J. Love and R.~W. Heath~Jr., ``Limited feedback unitary precoding for
  orthogonal space-time block codes,'' {\em IEEE Trans. on Sig. Proc.},
  vol.~53, pp.~64--73, Jan. 2005.

\bibitem{LoveHeath:LimitedFeedbackPrecodSpatialMultiplex:05}
D.~J. Love and R.~W. Heath~Jr., ``Limited feedback unitary precoding for
  spatial multiplexing systems,'' {\em IEEE Trans. on Info. Theory}, vol.~51,
  pp.~1967--76, Aug. 2005.

\bibitem{Gesbert:ShiftMIMOParadigm:2007}
D.~Gesbert, M.~Kountouris, R.~W. Heath, C.-B. Chae, and T.~Salzer, ``Shifting
  the {MIMO} paradigm,'' {\em IEEE Signal Proc. Magazine}, vol.~24, pp.~36--46,
  Sept. 2007.

\bibitem{Banister03}
B.~C. Banister and J.~R. Zeidler, ``Feedback assisted transmission subspace
  tracking for {MIMO} systems,'' {\em IEEE Journal on Sel. Areas in
  Communications}, vol.~21, no.~3, pp.~452--63, 2003.

\bibitem{SimonLeus:FbReductionSMLinearPrecod:2007}
C.~Simon and G.~Leus, ``Feedback reduction for spatial multiplexing with linear
  precoding,'' in {\em Proc., IEEE Int. Conf. Acoust., Speech and Sig. Proc.},
  vol.~3, Apr. 2007.

\bibitem{Huang:MarkovModelLimitFb:ICASSP06}
K.~Huang, B.~Mondal, R.~W. Heath, Jr., and J.~G. Andrews, ``{Markov} models for
  multi-antenna limited feedback systems,'' in {\em Proc., IEEE Int. Conf.
  Acoust., Speech and Sig. Proc.}, pp.~IV--9--IV--12, May 2006.

\bibitem{GallagerBook:InfoTheoReliableComm:68}
R.~Gallager, {\em Information Theory and Reliable Communication}.
\newblock Wiley, 1968.

\bibitem{3GPP-LTE}
``{3GPP TR} 25.814: Physical layer aspects for evolved universal terrestrial
  radio access (release 7),'' June 2006.

\bibitem{IEEE802-16e}
``{IEEE} 802.16e amendment: Physical and medium access control layers for
  combined fixed and mobile operation in licensed bands,'' {\em IEEE Standard
  802.16}, 2005.

\bibitem{Isu:FiniteRateFbCorrMISOEstmErrFbDly:2007}
Y.~Isukapalli and B.~D. Rao, ``Finite rate feedback for spatially and
  temporally correlated {MISO} channels in the presence of estimation errors
  and feedback delay,'' in {\em Proc., IEEE Globecom}, pp.~2791--2795, Nov.
  2007.

\bibitem{Ting:MarkovKronMIMO:2006}
S.~H. Ting, K.~Sakaguchi, and K.~Araki, ``A {Markov-Kronecker} model for
  analysis of closed-loop {MIMO} systems,'' {\em IEEE Commun. Lett.}, vol.~10,
  pp.~617--619, Aug. 2006.

\bibitem{Au:AdaptModMultiantTxChannelMeanFb:2004}
E.~Au, S.~Jin, M.~R. McKay, W.~Mow, X.~Gao, and I.~B. Collings, ``Analytical
  performance of {MIMO-SVD} systems in {Ricean} fading channels with channel
  estimation error and feedback delay,'' {\em to appear in IEEE Trans. on
  Communications}, Mar. 2008.

\bibitem{Nguyen:CapMIMOFbDelay:2004}
H.~T. Nguyen, J.~B. Andersen, and G.~F. Pedersen, ``Capacity and performance of
  {MIMO} systems under the impact of feedback delay,'' in {\em Proc., IEEE
  PIMRC}, vol.~1, pp.~53--57, Sept. 2004.

\bibitem{DuLi:PerformLossFbDelayMIMO:2004}
J.~Du, Y.~Li, D.~Gu, A.~F. Molisch, and J.~Zhang, ``Estimation of performance
  loss due to delay in channel feedback in {MIMO} systems,'' in {\em Proc.,
  IEEE Veh. Technology Conf.}, vol.~3, pp.~1619--22, Sept. 2004.

\bibitem{KuoSmith:SelfInterfMIMOFbDelay:2006}
P.-H. Kuo, P.~J. Smith, L.~M. Garth, and M.~Shafi, ``Instantaneous signal and
  self-interference power of {MIMO} eigenmode transmission with feedback time
  delay,'' in {\em Proc., IEEE Intl. Conf. on Communications}, vol.~9,
  pp.~4143--48, June 2006.

\bibitem{Kobayashi:MIMOFbDelay:2006}
K.~Kobayashi, T.~Ohtsuki, and T.~Kaneko, ``{MIMO} systems in the presence of
  feedback delay,'' in {\em Proc., IEEE Intl. Conf. on Communications}, vol.~9,
  pp.~4102--06, June 2006.

\bibitem{DuelHallen:FadingChanPredictAdaptTx:2007}
A.~Duel-Hallen, ``Fading channel prediction for mobile radio adaptive
  transmission systems,'' {\em Proceedings of the IEEE}, vol.~95,
  pp.~2299--2313, Dec. 2007.

\bibitem{Daniels07:QuantizedFbReductionSMLinearPrecod:2007}
R.~C. Daniels, K.~Mandke, K.~Truong, S.~Nettles, and J.~R.~W.~Heath,
  ``Throughput and delay measurements of limited feedback beamforming for
  indoor wireless networks,'' {\em Preprint:
  http://users.ece.utexas.edu/$^\sim$rdaniels/}, 2008.

\bibitem{Wang:FinStatMarkovChan:95}
H.~Wang and N.~Moayeri, ``Finite-state {Markov} channel -- a useful model for
  radio communication channels,'' {\em IEEE Trans. on Veh. Technology},
  vol.~44, pp.~163--71, Feb. 1995.

\bibitem{Pimentel:FiniteMarkovModelCorrRicianFading:2004}
C.~Pimentel, T.~H. Falk, and L.~Lisboa, ``Finite-state {Markov} modeling of
  correlated {Rician}-fading channels,'' {\em IEEE Trans. on Veh. Technology},
  vol.~53, pp.~1491--1501, Sept. 2004.

\bibitem{Tan:MarkovRayleighFading:2000}
C.~C. Tan and N.~C. Beaulieu, ``On first-order {Markov} modeling for the
  {Rayleigh} fading channel,'' {\em IEEE Trans. on Communications}, vol.~48,
  pp.~2032--40, Dec. 2000.

\bibitem{Zhang:MarkovModelRayleighFading:1999}
Q.~Zhang and S.~A. Kassam, ``Finite-state {Markov} model for {Rayleigh} fading
  channels,'' {\em IEEE Trans. on Communications}, vol.~47, pp.~1688--92, Nov.
  1999.

\bibitem{Babich:GenMarkovModelFlatFading:2000}
F.~Babich, O.~E. Kelly, and G.~Lombardi, ``Generalized {Markov} modeling for
  flat fading,'' {\em IEEE Trans. on Communications}, vol.~48, pp.~547--551,
  Apr. 2000.

\bibitem{Babich1999}
F.~Babich, G.~Lombardi, and E.~Valentinuzzi, ``Variable order markov modelling
  for {LEO} mobile satellite channels,'' {\em Electronics Letters}, vol.~35,
  pp.~621--623, Apr. 1999.

\bibitem{Chen:TractWlssChan:1998}
A.~Chen and R.~R. Rao, ``On tractable wireless channel models,'' in {\em Proc.,
  IEEE PIMRC}, vol.~2, pp.~825--830, Sept. 1998.

\bibitem{Saadani:MarkovRayleighRake:2004}
A.~Saadani, P.~Gelpi, and P.~Tortelier, ``A variable-order {Markov} chain based
  model for {Rayleigh} fading and {RAKE} receiver,'' {\em IEEE Signal
  Processing Letters}, vol.~11, pp.~356--358, Mar. 2004.

\bibitem{Guan:FSMCCorrFading:1999}
Y.~L. Guan and L.~F. Turner, ``Generalised {FSMC} model for radio channels with
  correlated fading,'' in {\em {IEE} Proc. Comm.}, vol.~146, pp.~133--137, Apr.
  1999.

\bibitem{Wang:VerifyFirstOrderMarkovRayleigh:1996}
H.~Wang and P.~Chang, ``On verifying the first-order {Markovian} assumption for
  a {Rayleigh} fading channel model,'' {\em IEEE Trans. on Veh. Technology},
  vol.~45, pp.~353--357, May 1996.

\bibitem{Sadeghi:CapAnalysisMarkovFlatFading:2005}
P.~Sadeghi and P.~Rapajic, ``Capacity analysis for finite-state markov mapping
  of flat-fading channels,'' {\em IEEE Trans. on Communications}, vol.~53,
  pp.~833--840, May 2005.

\bibitem{Bischl:PacketErrorNonInterleaveRayleigh:1995}
H.~Bischl and E.~Lutz, ``Packet error rate in the non-interleaved {Rayleigh}
  channel,'' {\em IEEE Trans. on Communications}, vol.~43, no.~234,
  pp.~1375--82, 1995.

\bibitem{Hueda:InfoTheoTestMarkovBlockErrors:2005}
M.~R. Hueda and C.~E. Rodriguez, ``A new information theoretic test of the
  {Markov} property of block errors in fading channels,'' {\em IEEE Trans. on
  Veh. Technology}, vol.~54, pp.~425--434, Mar. 2005.

\bibitem{Galluccio2005}
L.~Galluccio, F.~Licandro, G.~Morabito, and G.~Schembra, ``An analytical
  framework for the design of intelligent algorithms for adaptive-rate {MPEG}
  video encoding in next-generation time-varying wireless networks,'' {\em IEEE
  Journal on Sel. Areas in Communications}, vol.~23, pp.~369--384, Feb. 2005.

\bibitem{Turin:MAPDecodErrorBurst:2001}
W.~Turin, ``{MAP} symbol decoding in channels with error bursts,'' {\em IEEE
  Trans. on Info. Theory}, vol.~47, pp.~1832--38, July 2001.

\bibitem{Razavilar:OptimRateDelayControlFading:2002}
J.~Razavilar, K.~J.~R. Liu, and S.~I. Marcus, ``Jointly optimized bit-rate or
  delay control policy for wireless packet networks with fading channels,''
  {\em IEEE Trans. on Communications}, vol.~50, pp.~484--494, Mar. 2002.

\bibitem{Buche:ControlMobileCommTimeVarchan:2002}
R.~Buche and H.~J. Kushner, ``Control of mobile communications with
  time-varying channels in heavy traffic,'' {\em IEEE Trans. on Automatic
  Control}, vol.~47, pp.~992--1003, June 2002.

\bibitem{Yun:MarkovErrARQFading:2005}
J.~Yun and M.~Kavehrad, ``Markov error structure for throughput analysis of
  adaptive modulation systems combined with {ARQ} over correlated fading
  channels,'' {\em IEEE Trans. on Veh. Technology}, vol.~54, pp.~235--245, Jan.
  2005.

\bibitem{LiGoldsmith:MLDetectMarkov:2002}
L.~Li and A.~J. Goldsmith, ``Low-complexity maximum-likelihood detection of
  coded signals sent over finite-state {Markov} channels,'' {\em IEEE Trans. on
  Communications}, vol.~50, pp.~524--531, Apr. 2002.

\bibitem{Ivanis:AdpMIMOMRCImpCSIMarkov:2007}
P.~Ivanis, D.~Drajic, and B.~Vucetic, ``Performance evaluation of adaptive
  {MIMO}-{MRC} systems with imperfect {CSI} by a {Markov} model,'' in {\em
  Proc., IEEE Veh. Technology Conf.}, pp.~1496--1500, Apr. 2007.

\bibitem{Kuo:MarkovMIMOCondNumbAntSel:2007}
P.-H. Kuo, P.~J. Smith, and L.~M. Garth, ``A {Markov} model for {MIMO} channel
  condition number with application to dual-mode antenna selection,'' in {\em
  Proc., IEEE Veh. Technology Conf.}, pp.~471--475, Apr. 2007.

\bibitem{Abdi:LCRAvFadMIMOFading:2003}
A.~Abdi, C.~Gao, and A.~M. Haimovich, ``Level crossing rate and average fade
  duration in {MIMO} mobile fading channels,'' in {\em Proc., IEEE Veh.
  Technology Conf.}, vol.~5, pp.~3164--68, Oct. 2003.

\bibitem{TangHeath:OppFbDLMuDiv:2005}
T.~Tang and R.~W. Heath, Jr., ``Opportunistic feedback for downlink multiuser
  diversity,'' {\em IEEE Commun. Lett.}, vol.~9, pp.~948--950, Oct. 2005.

\bibitem{Huang:EffFbDelay:Globecom06}
K.~Huang, B.~Mondal, R.~W. Heath, Jr., and J.~G. Andrews, ``Effect of feedback
  delay on multi-antenna limited feedback for temporally correlated channel,''
  in {\em Proc., IEEE Globecom}, Nov. 2006.

\bibitem{Huang:LimitedFeedbackCompression:Globecom06}
K.~Huang, B.~Mondal, R.~W. Heath, Jr., and J.~G. Andrews, ``Multi-antenna
  limited feedback for temporally correlated channel: Feedback compression,''
  in {\em Proc., IEEE Globecom}, Nov. 2006.

\bibitem{GallagerBook:StochasticProcs:95}
R.~Gallager, {\em Discrete Stochastic Processes}.
\newblock Springer, 1995.

\bibitem{BremaudBook}
P.~Bremaud, {\em Markov Chains}.
\newblock Springer, 1999.

\bibitem{Ozarow:InfoTheoCellularMobile:1994}
L.~H. Ozarow, S.~Shamai, and A.~D. Wyner, ``Information theoretic
  considerations for cellular mobile radio,'' {\em IEEE Trans. on Veh.
  Technology}, vol.~43, pp.~359--378, May 1994.

\bibitem{PaulrajBook}
A.~Paulraj, R.~Nabar, and D.~Gore, {\em Introduction to Space-Time Wireless
  Communications}.
\newblock Cambridge, UK: Cambridge University Press, 2003.

\bibitem{Fill91:EigenvalBndsNonrevMarkovChain:91}
J.~A. Fill, ``Eigenvalue bounds on convergence to stationarity for
  non-reversible markov chain, with an application to the exclusion process,''
  {\em The annals of applied probability}, vol.~1, pp.~62--87, Jan. 1991.

\bibitem{AndrewsBook:WiMax:07}
J.~G. Andrews, A.~Ghosh, and R.~Muhamed, {\em Fundamentals of {WiMAX}:
  Understanding Broadband Wireless Networking}.
\newblock Prentice Hall, 2007.

\end{thebibliography}
